\documentclass[twocolumn,floats,amsfonts,amssymb,unsortedaddress,floatfix]{revtex4-1}
\input epsf
\usepackage{bm}
\usepackage{amsmath}
\usepackage{graphicx}
\usepackage{xcolor}
\usepackage{ulem}
\usepackage{lipsum}

\usepackage{hyperref}
\hypersetup{
    colorlinks=true, 
    linktoc=all,        
    linkcolor=black,  
    citecolor=blue
}

\begin{document}

\def\Bbb{\mathbb}

\title[Thermomechanics of DNA]{Thermomechanical Stability and Mechanochemical Response of DNA:\\ a Minimal Mesoscale Model}

\author{Cristiano Nisoli and A. R. Bishop}

\address{Theoretical Division,  Los Alamos National Laboratory, Los Alamos, NM, 87545, USA}
\email{cristiano@lanl.gov, cristiano.nisoli@gmail.com}
\begin{abstract}
{We show that a  mesoscale model, with a minimal number of parameters, can well describe  the  thermomechanical and mechanochemical behavior of homogeneous DNA  at thermal equilibrium under tension and torque. We predict critical temperatures for denaturation under torque and stretch, phase diagrams for stable DNA, probe/response profiles under mechanical loads, and the density of dsDNA as a function of stretch and twist. We compare our predictions with available single molecule manipulation experiments and find strong agreement. In particular we elucidate the difference between angularly constrained and unconstrained overstretching. We propose that the smoothness of the angularly constrained overstreching transition is a consequence of the molecule being  in the vicinity of criticality for a broad range of values of applied tension.}
\end{abstract}


\maketitle

\tableofcontents

\section*{Introduction}

DNA and its functions are recognized to be at the basis of the  nano-machinery of life \cite{Watson}. Yet, the nucleic acid macromolecule  is also in itself a very interesting and sophisticated nanomechanical object which can be exploited in material science. Spun almost as a thread it has been the basis for novel  materials~\cite{Seeman2, Seeman,Winfree, Yan}, such as DNA origamis~\cite{Pinheiro,Fu}, DNA  nano particles and DNA coated colloids~\cite{McFarlane,Nykypanchuk,DiMichele,Rogers}. Additionally DNA-carbon nanotube hybrids~\cite{Williams, Couet, Zheng, Zheng2}, have been the subject of much recent experimental  and numerical research~\cite{Williams, Couet, Zheng, Zheng2, Johnson3, Johnson, Gigliotti, Johnson2, Manohar, Yarotski, Kilina} as promising candidates for  nanotechnological applications in bio-molecular and chemical sensing, drug delivery~\cite{Williams, Gannon} and dispersion/patterning of carbon nanotubes~\cite{Zheng, Zheng2, Gigliotti}. 

Much of DNA's specificity both as a biological molecule and as a building block for novel materials comes from the interplay between the strong covalent bonds of the backbone and weak hydrogen interactions between bases. These latter are sensitive to thermal fluctuations---which indeed can lead to thermally induced denaturation at low ($\sim 70~^o$C) temperature. From a physics perspective, the weak inter base interactions dictates structure and symmetry, and their interplay with both mechanical fields and temperature  make DNA highly and non-linearly responsive, thus rendering a purely mechanical/energetic description insufficient. 

Because of these subtle thermal and mechanical couplings, DNA lends itself naturally to a thermomechanical analysis---to borrow an expression from material engineers---i.e. an analysis of its combined stability and average probe/response to both mechanical loads and temperature. Similarly, its non-linear load-induced response represents an interesting case of mechanochemistry~\cite{Hickenboth}. While this is critical for use of DNA as a building block in nanotechnology and possibly in bio-inspired self-healing materials, understanding double helix thermomechanics  also illuminates biology, where, e.g.,  enzymes involved in replication and repair are viewed as powerful molecular motors~\cite{Ma}. Although in biological applications such a level of description presents the limit of neglecting the fine structure given by base sequence~\cite{Sulc}---DNA's ultimate specificity---it never the less affords a baseline on which to build a more faithful analysis. Furthermore, it provides a rather faithful description of single molecule manipulation results, where the fluctuations in  strength of the base bond can be safely averaged away.

The direct single molecule manipulation~\cite{Bust, Strick,Bryant, Bust2, Mameren, Smith96} that has revolutionized our understanding of key aspects of DNA, might in fact be considered an experimental thermomechanical/mechanochemical analysis at the nanoscale. A few micrometer long DNA can be attached to beads and subjected to  applied tension and torque, possibly in varying concentration of salinity and at different temperatures, to reveal new couplings and transitions between different structures whose nature and forms, however, are still speculative. Given the length of the polymer employed, the fact that measured quantities (supercoiling, elongation) are all averages, and the relatively small---at least from a physicist's point of view---difference in interaction between different base pairs, one can assume that many of these results apply to an ideal, homogeneous DNA.

We are interested in the  study of the interplay between loads, temperature, and base bonds,  rather than on the complex topologies of DNA writhing, and thus we concentrate in particular on DNA kept straight by enough tension as to neglect formation of plectonemes:  from a few pico-Newtons (pN) up to $10^2$ pN. In this regime, experiments show sharp transitions at positive and negative torques~\cite{Bryant}; and several  years before those observations an ``overstretching transition'' had already been observed in DNA under tension of $\simeq 60-70$ pN~\cite{Bust, Bryant, LŽger}. All of these results have lead to tentative tension--torque phase diagrams for the stability of B--DNA~\cite{Bryant}, as well as to various phenomenological theories, some of which, highly parametrized, were proposed to explain these effects~\cite{LŽger,Sarkar,Marko,Fye, Bouchiat, Manghi, Nisoli1, Palmeri}. In particular, the robustness of the Peyrard-Bishop-Dauxois approach (PBD)~\cite{PB,PBD, Dauxois} has been  corroborated by modeling of Cocco and Barbi~\cite{Cocco, Barbi}: they incorporated torque and successfully reproduced denaturation by unwinding. However, these recent studies either do not include tension, or do not explain denaturation at overwinding, and do not provide  phase diagrams in the tension-torque plane. Also, although much simpler than the molecular structure they describe, their complexity cannot offer simple analytic equations to more easily guide experiments.

We show  here that the problem  is well suited to a minimal modeling that subsumes the tremendous complexity  of the DNA macromolecule under a very few relevant interactions and symmetries, providing average expectation values in a statistical mechanics fashion. The aim is to gain insight on what are in fact these relevant interactions and symmetries, and how to describe them. 

Building on a framework  announced in a recent letter~\cite{Nisoli11}, we offer an intuitive modeling of a long strand of DNA as a nano-material held under sufficient tension to avoid structural defects such as plectonemes. With respect to our previous work this treatment is self contained. We have included  the effect of bending fluctuations, and  in addition to the phase diagrams for stability we produce  probe/response profiles under applied field. We compare our analytical predictions with experimental results and find strong agreement both for phase diagrams and for non-linear responses. We then relate  responses to the density of open bases, and  correctly reproduce the linear behaviors reported in recent experimental results which combined force spectroscopy with fluorescent methods~\cite{Mameren}.  

This article is  organized as follows: in Section II ({\it Mechanics}) we detail the purely energetic part of the model, which includes torsional and bending deformations. In Section III ({\it Thermodynamics}) we perform the statistical mechanics treatment for the thermal fluctuations, and we show how our model can be reduced to a simpler Peyrard-Bishop (PB) model~\cite{PB} in which the effect of a mechanical field is folded back into a redefinition of the energy for the bases bond. This is of some practical importance, since there is widespread competence in numerical solutions of the PB model via numerical methods based on transfer matrix techniques. However, all the results we show here are purely analytical. In Section IV ({\it Thermomechanical Stability}) we solve the model in the context of the continuum limit and we propose phase diagrams for stability of B-DNA, whereas as well as profiles of probe/response under mechanical loads areexplained in Section V ({\it Non-Linear Mechanochemical Response}). There we compare our predictions with experimental results and we address the issue of the nature of the transition in our model, which quite faithfully reflect the transition in DNA as reported by recent experimental results~\cite{Bryant, Mameren}. 

\begin{figure*}[t!]   
\center
\includegraphics[width=2.9 in]{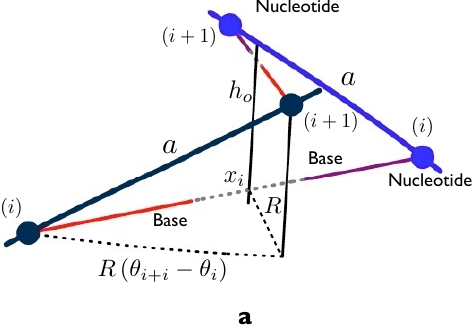}
\includegraphics[width=1. in]{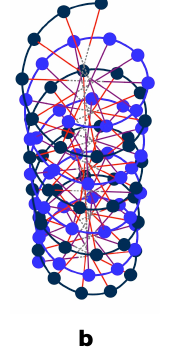}\hspace{10 mm}\includegraphics[width=1. in]{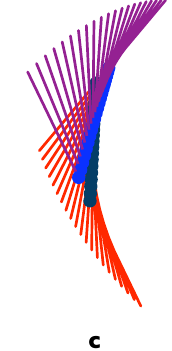}
\caption{(a) Schematics of our DNA model:  $i$ labels nucleotides separated by a distance $a$  along the DNA backbone,  $\theta_{i}$ is their angular coordinate, $x_i$ is the length of the $i^{th}$ base's bond. (b) A closed portion of the double helix. (c) We assume that in a region where base pairs are open, the two backbones twist with bases pointing outwards. Reprinted with permission from ref~\cite{Nisoli11}.}
\label{draw}
\end{figure*}

\section{Mechanics}

As motivated in the introduction, we assume that the DNA is homogeneous, or that the base dependence is averaged away in the thermodynamic limit of a long strand. Only average quantities, such as the average base bond strength, are relevant. In this section we develop the purely energetic model, in the following we will add the effect of thermal fluctuations.

\subsection{Free homogeneous DNA}

Our model has three sets of degrees of freedom:  $x_i$ is the length of the $i^{th}$ base's bond, whereas
\begin{equation}
\omega_i=(\theta_{i+1}-\theta_{i-1})/2 -\Omega
\end{equation}
 is the angular shift between nucleotides along the backbone, where $\theta_{i}$ is their angular coordinate; $i$ labels nucleotides separated by a distance $a$  along the DNA backbone (Fig. 1). $\Omega$ denotes the natural rotation ($\simeq2\pi/10$) per base pair of the B-DNA helix in equilibrium, and therefore $\omega_i$ describes deviations from equilibrium.   As in the Cocco--Barbi models~\cite{Cocco, Barbi},  the two strands of DNA are assumed symmetrical, with a fixed rigid center line, and infinitely long. A third degree of freedom, describing bending, will be introduced later in a perturbative fashion. For the moment we consider the molecule as straight.

The potential energy of of the system is
\begin{equation}
 E=a \sum_i F_i
 \label{e}
 \end{equation}
with 
\begin{equation}
F_i=\frac{k}{2} \frac{{\Delta x_i}^2}{a^2} +\frac{\nu}{2} \left(\omega_i+\Omega\right)^2 - \chi(x_i)V(\omega_i).
 \label{E}
\end{equation}
The first term in~(\ref{E}) is a stacking potential ($\Delta x_i=x_{i+1}-x_{i}$), which we  keep harmonic, as in the Peyrard-Bishop (PD) model~\cite{PB}. The second is an an elastic term which restores the $\theta_{i+1}=\theta_{i} $ (or equivalently $\omega_i=-\Omega$) flat angular configuration for the open strands. The third is a square well potential for the hydrogen bond between bases: $\chi(x)$ is a step function defined as 1 for $0\le x \le x^c$ and 0 for $x^c <x$, where $x^c$ is a length associated with the hydrogen bond. 

The depth of the square potential $V(\omega_i)$ represents the average strength of the base-base bonds, {
which includes the hydrogen bond but also effects of  any other interaction between backbones, such us electrostatic repulsion~\cite{DiMichele2}. Not knowing the specific and  complex form of the interaction,  we have subsumed all its information in a single quantity, $V$, depth of a square well, to avoid assumptions and parameter proliferation. Indeed, as we will discuss below, the existence of a transition follows from any interaction that can be described as a well on an half-line, and thus we choose a simple one.
(We will comment later on how change in ionic strength can affect such quantity.)
}
In our model, it depends   on the angle $\omega_i$: without this dependence and without the second term, the energy of (\ref{E}) would simply correspond to the PB model, or, with a non-linear first term, to the PBD model~\cite{PB,PBD}.  

The dependence of $V$ on $\omega_i$  follows, in the real molecule, from  a complex combination of hydrophobic, $\pi$--$\pi$, and dipolar interactions. It  reflects that  the bonding of opposite bases is responsible for DNA helicity. We can completely ignore the complexity of these interactions by  noting the following: first  $V(\omega)$ cannot be symmetric. If it were, DNA would not have a pitch. If we expand $V(\omega)$, it must be $V'(0)=\nu \Omega$, since the joined double strands of DNA, for which $\chi=1$ in (\ref{E}), must be in equilibrium at $\omega_i=0$.  Also, we must keep the expansion at second order, to recover the correct torsional rigidity. Indeed  $\mu=-V''(0)$ must be positive: Eq. (\ref{E}) shows that for a closed portion of DNA   $\nu+\mu$ is the purely mechanical torsional rigidity of the joined double helix. Since $\nu$ is the torsional rigidity of the much softer open portions of DNA ($\chi=0$) we have therefore $\mu=-V''(0) \gg \nu>0$ (we shall see, when fitting experimental data, that $\mu/\nu\sim 10^2$). With this in mind, we can now truncate  $V(\omega)$ to the second order as
\begin{equation}
V(\omega)\simeq V_0+\nu\Omega \omega-\frac{1}{2}\mu \omega^2,
\label{V}
\end{equation}
where $V_0$ is the average strength of the base-base bond in equilibrium, $\omega_i=0$.

Equation~(\ref{V}) correctly reflects the  the strengthening of base bonds under overwinding, and weakening under  unwinding, already presaging the stabilizing (destabilizing) effect of a positive (negative) torque, something well known experimentally~\cite{Bryant}. However, the quadratic term in (\ref{V}) also implies that large enough overwinding  can eventually  melt DNA, in agreement with intuition as well as experimental evidence~\cite{Bryant}. 

One last note on $V(\omega)$:  truncation of the expansion at  second order is sufficient to study the in stability of B-DNA toward denaturation by strand separation. However introducing a third (and thus necessarily also a fourth) term might prove useful in the future to investigate other structural transitions in the pitch, for instance between A-, B-,  and Z-DNA, as well as to study the  correlation of torsional fluctuations. 


\subsection{Mechanical loads}

Let us assume that the system is held under a tension $f$ and under torque $\Gamma$.  The torque $\Gamma$  is  easily incorporated in (\ref{E}) via a term -$\sum_i \tau \omega_i a$.   (For dimensional convenience we keep all the constants as forces;  we thus introduce $\tau=\Gamma/a$  and for simplicity we refer to both $\Gamma$ and $\tau$ as torques in the following, even though $\tau$ is a force.)

Tension however is more subtle: closed and open portions of DNA provide different tensile responses.
The contribution of tension $f$  to the energy is $-f \Delta h$ and the total stretch $\Delta h$ is  readily obtained as 
\begin{equation}
\Delta h=\sum_i [1-\chi(x_i)] \Delta h_i^{o} + \sum_i\chi(x_i)\Delta h_i^{c}
\label{stretch}
\end{equation}
and has  contributions from both closed ($\Delta h_i^c$) and open ($\Delta h_i^o$) DNA sections. We consider the backbone as effectively inextensible and then $\Delta h$ arises only from winding/unwinding (and also from bending fluctuations---which we will treat later). 

\begin{figure}[t!!!!]   
\center
\includegraphics[width=3. in]{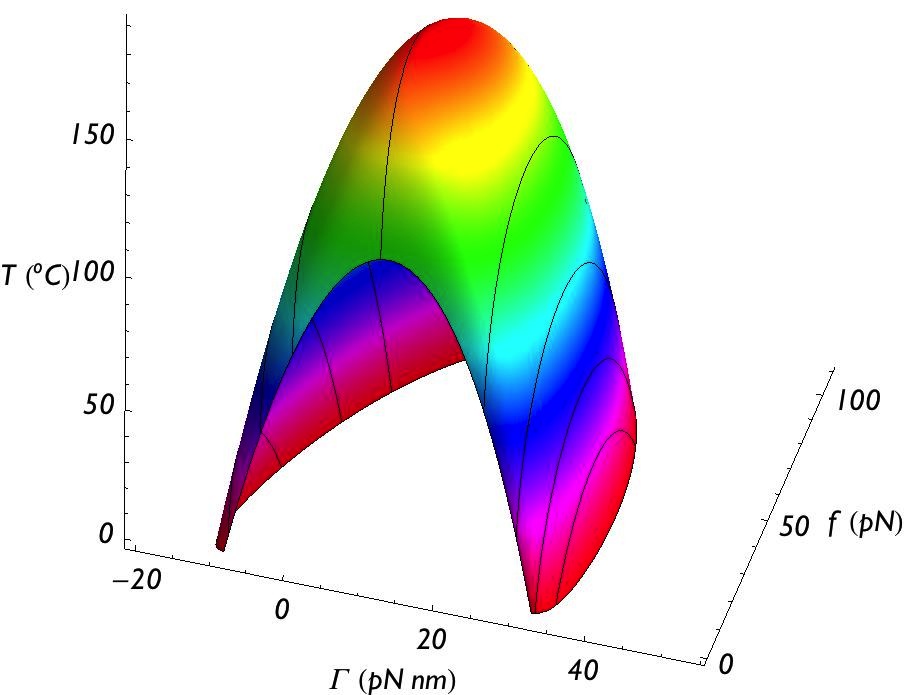}
\caption{Predicted critical surface for denaturation of B--DNA in ($T$), torque ($\Gamma$) and tension ($f$) obtained from (\ref{r}). Solid lines on the surface correspond to critical lines at fixed torque $\Gamma$ plotted in Fig.~\ref{Tf}. The existence of a maximum at non zero tension points to the stabilizing effect of tension in suppressing flexural modes. This effect strengthens  the effective base-base potential by a term $ T\sqrt{f}/\sqrt{\lambda^o}$ which, at small $f$, counteracts the tensile energy gain for base opening $-f v$, see (\ref{W0}).}
\label{Surface}
\end{figure}

Let us compute the stretch $\Delta h$ as a function of changes in the winding angle.  When DNA's two strands are joined,  the vertical distance between nucleotides as a function of $\omega_i$ is simply ${h^c}^2=a^2-R^2(\Omega+\omega_i)^2$ (Fig. 1). We are interested in $\Delta h^c=h^c-h_0$, where $h_0^2=a^2-R^2\Omega^2 $ is the vertical distance between nucleotides at equilibrium (Fig. 1). Expanding in $\omega_i$ and truncating at the second order we obtain the stretch due to changes in angle for a closed portion of DNA as
\begin{equation}
\frac{\Delta h^c}{a}=-\frac{R^2\Omega}{h_0 a}  \omega_i -  \frac{R^2a}{2h_0^3} \omega_i^2.
\label{stretchc}
\end{equation}
Gore and  collaborators~\cite{Bust2} have reported an anomalous overwinding of DNA under stretch, which has been confirmed by further experiments. Although the phenomenon has noot been fully explained to this day, most analysis attribute it to a positive coupling between elongation and overwinding in the purely mechanic energy of ssDNA. As the negative sign in the first term of (\ref{stretchc}) show, such coupling is rather counterintuitive in an helical geometry. We could include such coupling here  by simply replacing the quantity ${R^2\Omega}/{h_0 a}$ with a negative parameter to be fitted later, as it is often done~\cite{Bust2,Gross}. However the structural origin of such in ssDNA  topology would still be mysterious or speculative. Furthermore, the origin of the anomalous overwinding might be thermodynamical, rather than structural: as tension suppresses destabilizing bending modes, it might increase the fraction of ssDNA, leading to an overall supercoling. We therefore choose not to introduce such coupling artificially  and without a solid physical grounding, however we discuss it at the end of section IV.

To explain the coupling between stretch and twist when the strands are {\it open} further assumptions on structure are necessary.  We assume that, when open and under torque, the two backbones twist as two ropes, with bases pointing outwards as in P-DNA (Fig.~1). This assumption is corroborated by measures with fluorescent molecules that bind to DNA's exposed bases~\cite{Mameren}. This configuration is also an helix of a certain radius $r<R$, which can be interpreted as an effective radius  for the backbone. {This is ``effective'' in more than one sense:  firstly one does not know exactly how to define the radius of a backbone, which is  not a cylindrical object; secondly the distance of the twisting backbones is affected by mutual repulsion~\cite{DiMichele2} and thus by ionic strength.} We can now proceed as above, but  we must expand around $\omega_i=-\Omega$ which is the stable configuration in which  the two backbones are stretched parallel to each other with no winding. From standard trigonometry we obtain for $\Delta h^o=h^o-h_0$ 
\begin{equation}
\frac{\Delta h^o}{a}=1-\frac{h_0}{a}-\frac{r^2}{2a^2}(\Omega +\omega_i)^2.
\label{stretchopen}
\end{equation}
%
Note that in (\ref{stretchopen}), if $\omega_i=-\Omega$, then $\Delta h^o=a-h_0$ is the elongation due to pure base opening.

Equations (\ref{stretchc}) and (\ref{stretchopen}) would seem to doom the expression for the energy (\ref{E}) to a certain undesired mathematical uncleanness. However, all of the considerations of Section 2.1, which lead to the energy in  (\ref{E}), where based on symmetry alone, which is not changed by tension; only the parameters are. One thus expects  that  the effect of tension can be subsumed into a redefinition of the parameters introduced so far, resulting in an expression of the energy that has the same functional form as in (\ref{E}), in the new, tension-dependent parameters. That is indeed the case. 

It is useful to introduce the following quantities, renormalized by tension, 
\begin{eqnarray}
\tilde\mu&=&\mu+m f, \nonumber \\
\tilde\nu&=&\nu+n f, \nonumber \\ 
\tilde \Omega&=&\Omega-\frac{o f}{\tilde \mu +\tilde\nu},  \nonumber \\ 
 \tilde \omega&=& \omega +\frac{o f}{\tilde \mu +\tilde\nu}, \nonumber \\
\tilde V_0 &=&V_0-f v -\frac{\nu}{2} \Omega^2+\frac{\tilde \nu}{2} {\tilde \Omega}^2+\frac{\tilde \mu+\tilde \nu}{2}\left(\Omega-\tilde \Omega\right)^2,
\label{rin}   
\end{eqnarray}
where  $m=aR^2 h_0^{-3}-n$, $n=r^2a^{-2}$, $o=  \Omega R^2a^{-1} h_0^{-1}$, $v=1-h_0/a$ are dimensionless and purely geometrical parameters which, with the exception of $r$ the radius of rope-twisting of open backbones, can be obtained from the known structure of DNA. We take here $R\simeq10$ \AA~for the radius of the DNA molecule,  $h_0\simeq3.4$ \AA~for the elevation between consecutive nucleotides, $a\simeq 7$ \AA~ for their distance along the backbone, and $\Omega=2\pi/10$ for the rotation per base pair of DNA (Fig. 1): these are established geometrical values for B--DNA, but our formalism works, {\it mutatis mutandis}, for A-- and Z-- forms. Instead $r$, the effective radius around which the ssDNA filaments twist has to be fitted with experimental data (below). 

\begin{figure}[t!]   
\center
\includegraphics[width=3. in]{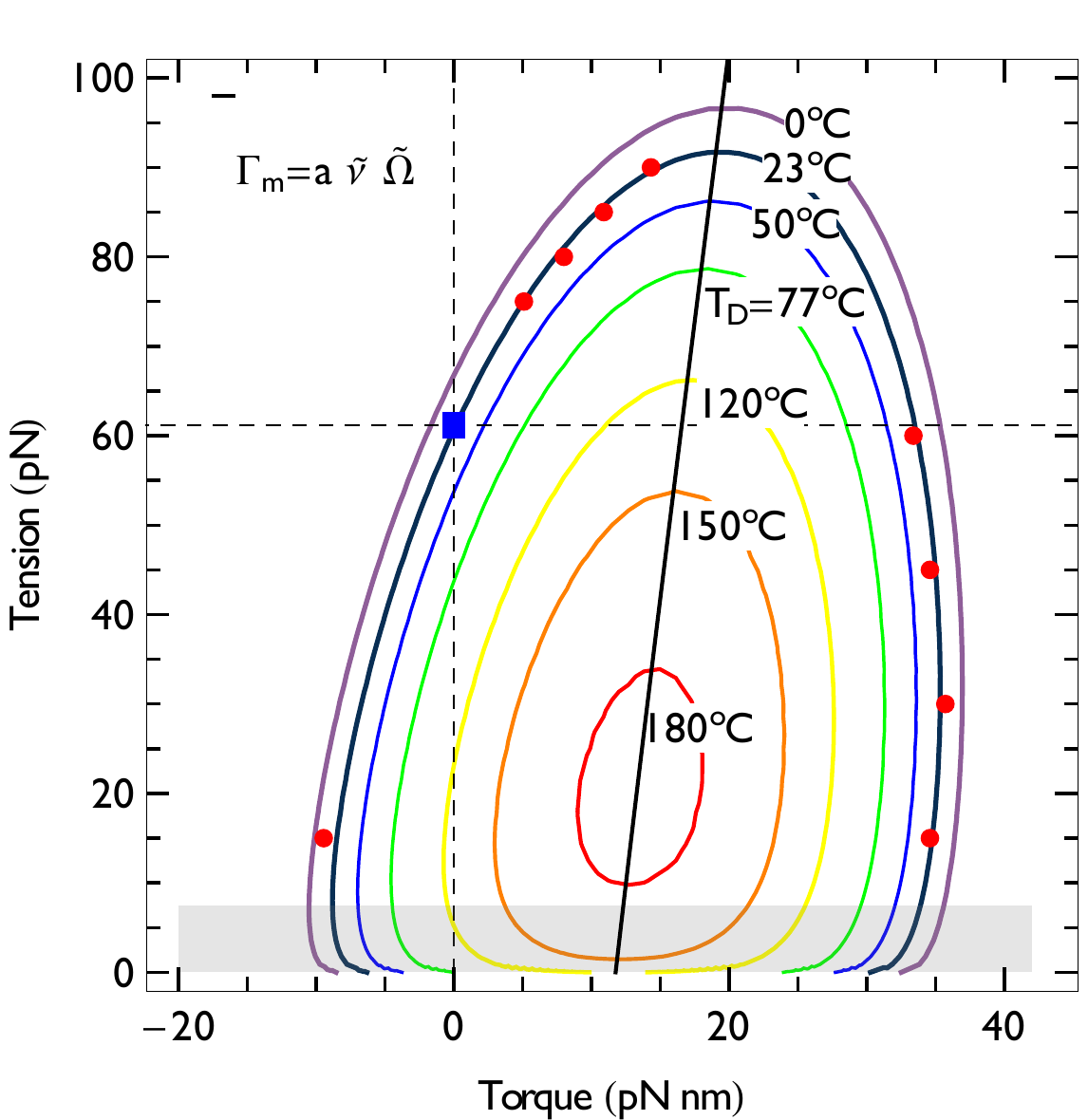}
\caption{Predicted critical lines for denaturation at different temperatures. The region enclosed by each line corresponds to stable B--DNA at corresponding temperature. Points are experimental data from single molecule manipulation~\cite{Bryant}, square corresponds to the well known over-stretching transition ($\Gamma=0$, $f=60$ pN). $T_D=350$ K $=77~^o$C, the denaturation temperature, is critical at zero external load. Yet but even at temperatures of $T_D$ or higher DNA can be stable in an interval of applied tension and torque. Solid (dotted) straight line indicates $\Gamma_m(f)$ given by (\ref{Gm}), the middle point between critical torques, which also corresponds to the highest critical temperature at given tension (dashed line in Fig~\ref{Tf}). At very high temperatures ($\ge 120~^o$C) DNA  is stable only under applied tension and positive torque. 
The translucent rectangle denotes a low tension region in which plectonemes can form and our approach does not apply.}
\label{ft}
\end{figure}

From $ \omega_i=\Delta\theta_i- \Omega$ and (\ref{rin}) follows
\begin{equation}
\tilde \omega_i=\Delta\theta_i-\tilde \Omega,
\label{omegatilde}
\end{equation}
which, as we will see below, suggests itself as the new natural angular variable when tension is present.

We can then write  the energy in presence of mechanical loads as
\begin{equation}
 \tilde E=a \sum_i \tilde F_i
 \label{etilde}
 \end{equation}
where  
\begin{equation}
\tilde F_i=F_i-f\Delta h/a-\tau \omega_i
 \label{etilde2}
 \end{equation}
and $\Delta h$ is given by  (\ref{stretch}).

Then, as anticipated, a rather tedious algebra shows that $\tilde F_i$  can be expressed in a quite compact way that has the same form as (\ref{E}) but in the new variable $\tilde \omega$ and new renormalized quantities reported in (\ref{rin}). From (\ref{stretchc}), (\ref{stretchopen}), and (\ref{rin}) we have
\begin{eqnarray}
\tilde F_i=&& \frac{k}{2} \frac{{\Delta x_i}^2}{a^2} +\frac{\tilde \nu}{2} \left(\tilde \omega_i+\tilde \Omega\right)^2  - \chi(x_i) \tilde V (\tilde\omega_i) -f v-\tau \tilde \omega_i \nonumber \\
\label{Enrin}
\end{eqnarray}
with, for $\tilde V$, the ``renormalized'' form of $V$ in (\ref{V})
\begin{eqnarray}
\tilde V= \tilde V_0+\tilde \nu \tilde \Omega \tilde \omega_i-\frac{\tilde\mu }{2} \tilde \omega_i^2.
\label{Vtilde}
\end{eqnarray}

Clearly $\tilde \Omega$ is the new rotation per base for closed portions of DNA under tension $f$. Indeed, when $\chi=1$ in (\ref{Enrin}) the energy (\ref{etilde}) is a  quadratic form in $\tilde \omega$ with minimum in $\tilde \omega=0$ or $\Delta \theta=\tilde \Omega$. Note that the by fitting $o$ as a negative number,  rather than by taking it from its definition above, one can obtains the anomalous stretch/twist coupling discussed above: then $\tilde \Omega$  increases with tension.

 Then $\tilde \omega=\Delta\theta-\tilde \Omega$ represents the angular deviation with respect to the tension-renormalized gain angle of closed portions. For open portions of DNA, i.e. when $\chi=0$, $\tilde F_i$ in (\ref{Enrin}) has a minimum at $\tilde \omega=-\tilde \Omega$ or $\Delta\theta=0$.

\begin{figure*}[t!!!!]   
\includegraphics[width=3 in]{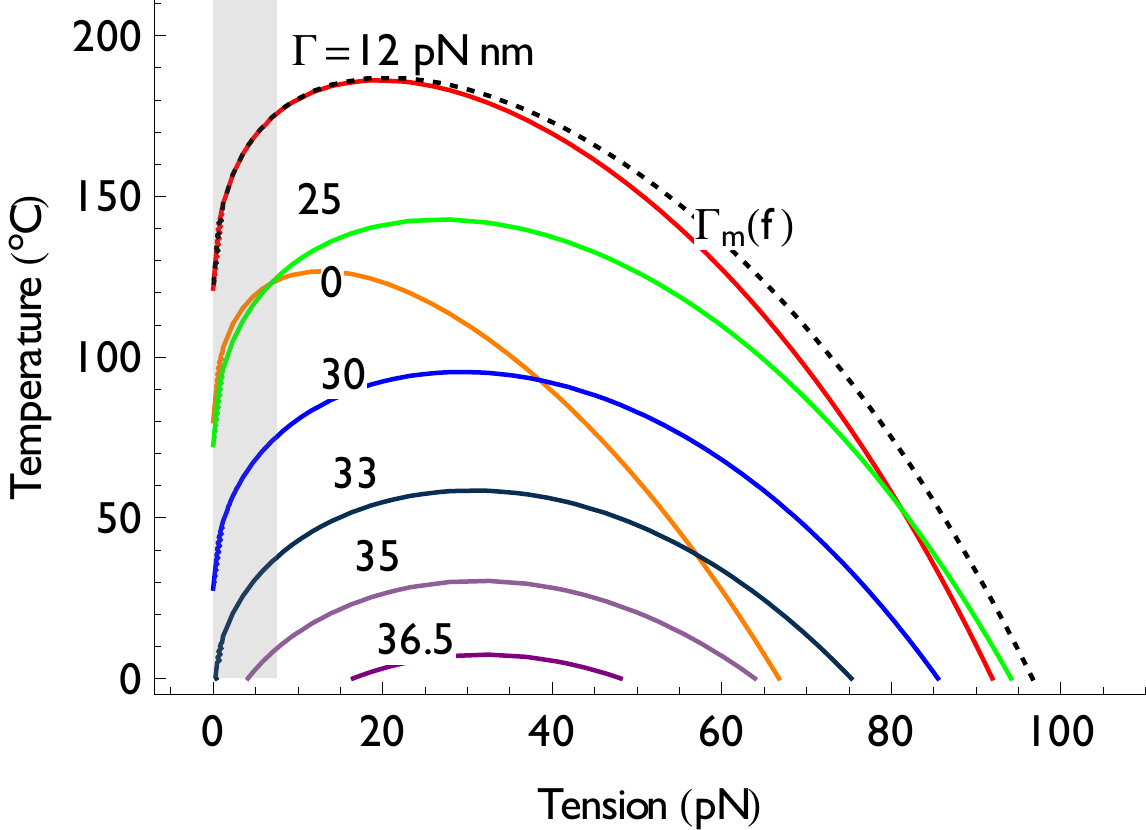}\includegraphics[width=3 in]{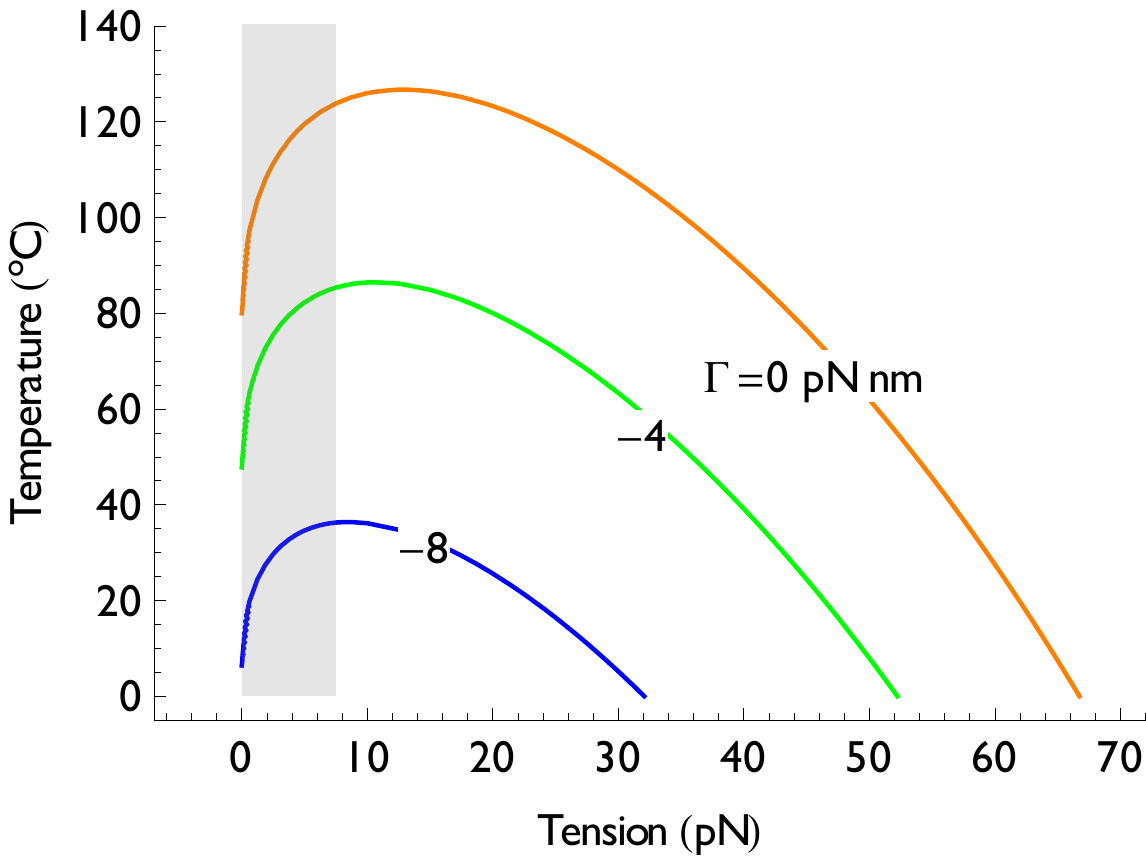}
\caption{Predicted critical temperature as a function of tension for DNA twisted under different torque. (Left: positive torque. Right: Negative torque.). Numbers on curves denote the applied torque in pN$\times$nm. The maximum critical temperature at any given tension (dotted curve) is reached at external torque $\Gamma_m(f)$ from (\ref{Gm}).  At large positive torque ($\ge 33$ pn$\times$nm)  DNA is only stable under a tension that contributes to the effective torque  (\ref{taueff}). Note how in general low tension  stabilizes DNA by suppressing flexural modes.  The translucent rectangle denotes a low tension region in which plectonemes can form and our approach does not apply.}
\label{Tf}
\end{figure*}

Similarly, (\ref{Vtilde}) and the first  line of  (\ref{Enrin}) have the same form as (\ref{V}) and (\ref{E}) with the replacement $\omega_i \to\tilde \omega_i$, $\Omega \to \tilde \Omega$, $\mu \to \tilde \mu$, $\nu \to \tilde \nu$, $V_0 \to \tilde V_0$. Therefore  $\tilde \mu$, $\tilde \nu$  from  (\ref{rin})  are the effective torsional rigidities for the system under applied tension $f$. 
The fourth term in (\ref{Enrin}) describes the remnant effect of the tension, besides renormalization of the angular variable and parameters: $-f v~Na$ is the work done by tension when opening bases ($v$, introduced above as $v=1-h_0/a$ is the relative elongation per base length due to base opening), whereas $-\tau \tilde \omega_i$ is the usual torque term. 

Finally, consider a closed portion of DNA, in angular equilibrium with $\omega_i=0$ for every $i$, and therefore no deviations from the helical angle $\Omega$. Then a (small) tension $f$ is applied, so that the resulting configuration corresponds to dsDNA in equilibrium at $\tilde\omega_i=0$, and therefore of new helical angle $\tilde \Omega$. Then the work done by the tension is  $f \Delta h=E_{|_{\omega=0}}-E_{|_{\tilde \omega=0}}$ (both energies computed at  $\chi=1$) which, from ({\ref{e}), ({\ref{etilde}), ({\ref{Enrin}), and (\ref{rin}), can be expressed  as
\begin{equation}
f\Delta h= \frac{\tilde \mu+\tilde \nu}{2}\left(\Omega-\tilde \Omega\right)^2,
\end{equation}
 intuitively the torsional elastic energy in the new helical angle and torsional rigidities. 
We have thus effectively subsumed the effects of tension, which are different on open and closed DNA and, as expressed by (\ref{stretch}), (\ref{stretchc}) and (\ref{stretchopen}), rather complicated, into a compact redefinition of the variables and parameters of the hamiltonian.

\subsection{Mechanical response}

In the absence of thermal fluctuations, our DNA model  is still  purely energetic. We can study the mechanical response of open and closed portions of DNA separately by considering $\tilde \omega_i=\tilde \omega$ as uniform yet different in the open and closed DNA cases. 

Response to applied fields can be computed by minimization of the energy in (\ref{etilde}). If we neglect bending, from $\partial \tilde E/\partial{\tilde \omega}=0$ we obtain 
\begin{eqnarray}
\tilde\omega&=&\frac{\tau+ \tilde \nu \tilde \Omega (\chi-1)}{\tilde \nu +\chi \tilde \mu},
\label{omegamin0}
\end{eqnarray}
where as usual $\chi=1$ ($\chi=0$) for closed (open) DNA or, more explicitly
\begin{eqnarray}
\tilde \omega^o&=&-\tilde \Omega +\frac{\tau}{\tilde \nu} \nonumber \\
\tilde \omega^c&=&\frac{\tau}{\tilde \mu+\tilde \nu}.
\label{omegamin}
\end{eqnarray}
From (\ref{omegamin}) and (\ref{omegatilde}) we find $\Delta \theta$ in closed and open portions of DNA:
\begin{eqnarray}
\Delta \theta^c&=&\tilde \Omega+\tau/(\tilde \mu +\tilde \nu) \nonumber \\
\Delta \theta^o&=&\tau/\tilde \nu .
\label{Deltathetac0}
\end{eqnarray}
While these results are certainly not surprising, they will be useful in the following, when we will express the response to fields under thermal fluctuations in terms of the purely energetic responses and the density of open/closed bases. 

It is worthwhile to make explicit the dependence on tension of the DNA gain angle. Substituting the values of  (\ref{rin}) into  (\ref{Deltathetac0}) we get, for closed portions of DNA (or for a DNA that is forced to be closed, such a methylated DNA),
\begin{equation}
\Delta \theta^c= \Omega+\frac{\tau_{\mathrm{eff}}} { \mu + \nu}
\label{Deltathetac}
\end{equation}
where we have introduced an effective tension-dependent torque 
\begin{equation}
\tau_{\mathrm{eff}}=\frac{\tau-o f}{1+f\frac{m+n}{\mu+\nu}}.
\label{taueff}
\end{equation}

Equation (\ref{taueff}) shows that at small tensions, or $f \ll (\mu+\nu)/(m+n)\sim40$ pN (we will see in the section devoted to fits of experimental data that $\mu+\nu\simeq 10^3$ pN), the tension $f$ exerts an effective unwinding torque per unit length equal to $\tau_{\mathrm{eff}}=\tau-of=\tau-\Omega f R^2 /ah_0$. As explained above, if one introduces the anomalous stretch/winding coupling, then $o$ is negative and tension induces a positive effective torque.  

A purely mechanical stretch can be obtained in a similar fashion. We will not report it here, as in any real application the effect of thermal bending modes on stretch cannot be neglected~\cite{Palmeri,Yan2}. 

\subsection{Perturbative effect of bending}

In addition to the degrees of freedom introduced so far, i.e. $\{x_i\}$ and $\{\omega_i\}$, we consider  here the bending of the system, which affects the change in length and is therefore coupled to tension. Also, bending  affects the strength of base bonds. It thus introduces a new coupling with tension which can potentially compete with torsion.  

We are concerned here only with small deviations from a straight line, since DNA has a long persistence length, and since we aim to describe single molecule manipulation experiments in which DNA is held (approximately) straight in a tension regime that excludes plectonemes. We thus introduce the two dimensional vector $\vec \psi_i$ which describes the deviation of the DNA chain from a straight line of the experimental apparatus, at the nucleotide $i$.  Then the DNA strand held under tension thermally fluctuates giving rise to bending modes, which in turn also  cause a contraction.

Since in the approximation of strong tension $f$ we exclude plectonemes or other structurally complex configurations, we can introduce the discretized curvature of DNA as $k_i=|\vec\psi_{i+1}+\vec\psi_{i-1}-2\vec\psi_i|/a^{2}$. Then, to include the effect of bending,  a term $a \sum_i \Delta \tilde F^{\psi}_i$ has to be added to the energy to obtain
\begin{equation}
 \tilde E^{\psi}=a \sum_i \tilde F_i+a \sum_i \Delta \tilde F^{\psi}_i,
 \end{equation}
where $\Delta \tilde F^{\psi}_i$ has the form
\begin{equation}
\Delta \tilde F^{\psi}_i=[1-\chi(x_i)]\frac{\lambda^o}{2}k_i^2 +\chi(x_i)\frac{\lambda^c}{2} k_i^2 +\frac{f}{2} |\Delta\vec\psi_i|^2.
\label{bending}
\end{equation}
The first two terms describe the elastic cost of curvature for open and closed portions of DNA ($\lambda^o$, and $\lambda^c$ are the corresponding bending rigidities, and clearly $\lambda^o \ll \lambda^c$). The third term reflects the coupling  with the applied tension and comes from $-f \Delta h^{\psi}_i$, where $\Delta h^{\psi}_i\simeq-|\Delta\vec\psi_i|^2/2$ is the local change in vertical coordinate due to the bending when $\Delta \vec \psi_i=(\vec\psi_{i+1}-\vec\psi_{i})/a$  is small. From (\ref{bending}) we see that  the strength of the base-bond in (\ref{Vtilde}) is further modified by an extra term
\begin{equation}
\Delta\tilde V^{\psi}=-(\lambda^c-\lambda^o)k_i^2/2.
\label{bending2}
\end{equation}
Thus bending effectively reduces the strength of the base-base bonds, by a term  
proportional to the square of the local curvature of dsDNA. 

\begin{figure}[t!]   
\center
\includegraphics[width=3 in]{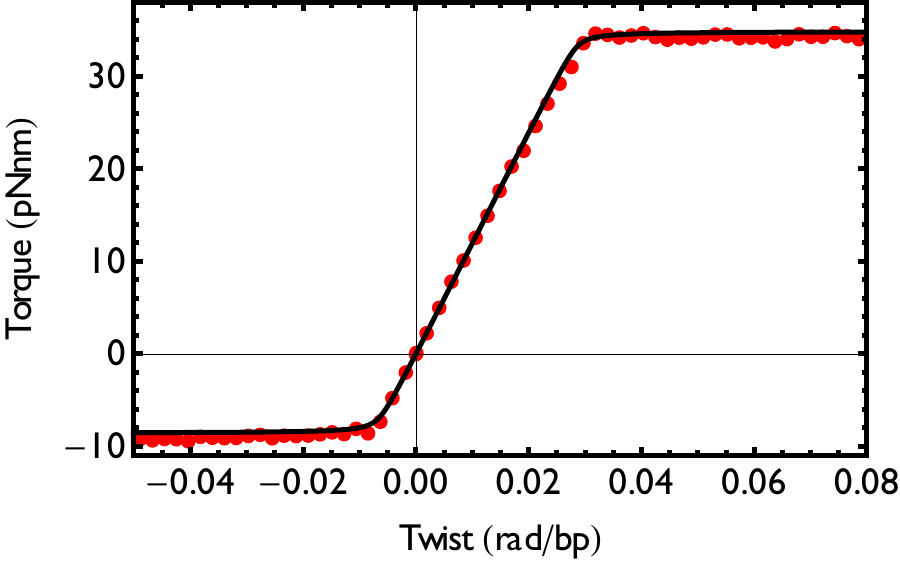}
\caption{Plot of applied torque vs. twist of a DNA strand; red dots corresponds to experimental data from Ref.~\cite{Bryant}. The straight line is our prediction based on  (\ref{omegaav4}).} 
\label{pappa}
\end{figure}

\section{Thermodynamics}

In this section we develop the field driven statistical mechanics of our model by summing over the thermal fluctuations in the base bonds, torsional, and flexural modes. We will show that the thermomechanics  of our model can be reduced to the thermodynamics  of a PB or PBD~\cite{PB, PBD} model in the sole $x_i$ variable, which describes the opening of bases. The effect of torsional and flexural modes is subsumed into an effective, field-dependent strength of the base bond. 

\subsection{Equivalent Peyrard-Bishop-Dauxois model}

Equilibrium thermodynamics is implemented by integration over all the configurations specified by the coordinates $\{x_i\}$, $\{\tilde \omega_i\}$  and $\{\psi_i\}$ to obtain the partition function
\begin{equation}
Z=\int \prod_i dx_i d\tilde\omega_i d\psi_i \exp{\left[-\beta a\left(\tilde F_i+\Delta \tilde F^{\psi}_i\right)\right]}
\label{Ztot}
\end{equation}
($\beta=1/k_B T$, where $k_B$ is the Boltzmann constant). The result is simple and we state it immediately (the derivation is discussed later):  both the angular and the flexural degrees of freedom can be integrated out, to reduce the partition function to an equivalent PB~\cite{PB} model, with square well potential, in the variables $\{x_i\}$, or 
\begin{equation}
Z(W,\Delta)=e^{-\beta N a\Delta} \int\prod_i dx_i e^{-\beta a\left[ \frac{k}{2} \frac{{\Delta x_i}^2}{a^2}- \chi(x_i)W \right]}.
\label{Z}
\end{equation}
The effect of tension and torque has been subsumed into the strength of a square, effective potential whose depth is given by
\begin{equation}
 W_{\tau,f,T}=W_{0,f,T}+ \tilde \Omega \tau -\frac{1}{2} \frac{\tilde \mu
}{(\tilde \nu+\tilde \mu)\tilde \nu} \tau^2.
\label{W}  
\end{equation}

Equation  (\ref{W})   incorporates explicitly the effect of the external torque $ \tau=\Gamma/a$. The effect of  tension $f$, is implicit in the tension-increased torsional rigidities $\tilde \mu$, $\tilde \nu$, in the pitch under tension $\tilde \Omega$ defined in~(\ref{rin}), and in $W_{0,f,T}$, the depth of the effective potential in the absence of external torque, which is given by
\begin{eqnarray}
W_{0,f,T}&=& V_0-\frac{\nu}{2} \Omega^2 +\frac{\tilde \nu+\tilde \mu}{2}\left(\Omega-\tilde\Omega\right)^2  -f v  +T\sqrt{f}\Delta \frac{1}{\sqrt{\lambda}} \nonumber \\
&-&\frac{T}{2a}\ln\frac{\tilde \nu+\tilde \mu}{\tilde \nu}-\frac{T}{a}\ln\frac{\lambda^c}{\lambda^o},
\label{W0}
\end{eqnarray}  
where $\Delta ({1}/{\sqrt{\lambda}})= 1/\sqrt{\lambda^o}- {1}/{\sqrt{\lambda^c}}>0$. Thus $W_{0,f,T}$ controls denaturation under tension or temperature alone.

While $W_{\tau,f,T}$ controls stability, the term 

\begin{widetext}
\begin{equation}
\Delta= -f v +T\sqrt{\frac{ {f}}{ {\lambda^o}}}+\tau \tilde \Omega  -\frac{1}{2\tilde \nu} \tau^2-\frac{T}{2a}\ln\frac{2\pi T}{a \tilde \nu}-\frac{T}{a}\ln\frac{2\pi T a}{\lambda^o}
\label{Delta}
\end{equation}
\end{widetext}
in (\ref{Z}) is absent in the PB model and irrelevant for the phase diagram in our model.  It  must however be kept when computing the specific heat and in general probe/response quantities, such as the  average supercoiling per base $\langle \Delta \theta \rangle$, as we will see in the next section. 

We are now in familiar territory, as the dynamics and thermodynamics of the PB model has been studied extensively, with different potentials (see Ref.~\cite{Rapti} and references therein). Standard techniques to solve it include continuum limit (analytic) or more often numerical methods based on the application of transfer matrix to a (quasi) one-dimensional system.  
Since the integration over the angular and flexural configurations do not involve the stacking potential, the same considerations apply to a non-linear choice of the stacking potential, such as the one used in the PBD model~\cite{PBD}, which is known to better reproduce the sharpness of the transition, in the absence of loads and therefore at higher temperatures. 

However, we will see below that our model can capture the sharpness of the transition without added non-linearity in the stacking potential. Indeed, note that unlike in the PD model now  the  depth of the effective potential, even in absence of torque and tension, depends on temperature in a non trivial way. Temperature weakens the effective potential via terms that translate the disrupting effect of angular and flexural fluctuations. This, we will see, has consequences on the possibility to affect the sharpness of the transition within the model.

\subsection{Thermal integration of angular and flexural modes}

We now discuss how  (\ref{Z}), (\ref{W}), (\ref{W0}), (\ref{Delta}) are obtained. The integration over the angular variable is trivially gaussian, since  everything in $\tilde F_i$ is quadratic in $\tilde \omega_i$.  Then one immediately recognizes in the $\tau$-dependent terms of $W_{\tau,f,T}$ and $\Delta$ simply the $\tilde F_i$ computed on the minimal $\tilde \omega_i$ given by (\ref{omegamin}). In addition one must consider the ``equipartition'' terms coming from the integration of the quadratic fluctuations around the minimum: these are the first term in the second line of (\ref{W0}), and the fifth term in  (\ref{Delta}).

All the other terms  come from integration of the flexural modes, the integral over $d\psi_i$ in (\ref{Ztot}), which is carried out approximatively. First we take again a continuum limit, $i \to s$, $x_i\to x(s)$, and $\left(\vec \psi_{i+1}-\vec \psi_{i}\right)/a \to \vec \phi(s)$, $k_i\to \phi'(s)$, $a\sum_i\to \int ds$. Note that this approximation is always well justified and does not necessarily imply the adoption of the same limit in the resolution of the PB model of (\ref{Z}): indeed  the persistence length of DNA is notoriously much larger than $a$, even in the absence of straightening tension. 

From (\ref{bending}) the integration adds a factor
\begin{widetext}
\begin{equation}
{\cal I[}x(s)]\propto\int {\cal D}\vec \phi(s) \exp\Bigg\{{-\beta \int_0^N a \left[ \frac{\lambda(x(s))}{2}|\vec\phi'(s)|^2+\frac{f}{2}|\vec\phi(s)|^2\right]ds\Bigg\}}
\label{factor}
\end{equation}
\end{widetext}
to the integrand of (\ref{Ztot}), where, from (\ref{bending}),  $\lambda(x)=[1-\chi(x)]\lambda^o+\chi(x)\lambda^c$. 
Since $x(s)$ is fixed, the integral in (\ref{factor}) factorizes into many integrals corresponding to open or closed portions of lengths $\{L^{o,c}\}$, controlled by $x(s)$. Each of these integrals corresponds to a propagator in the imaginary time $L$ for an harmonic two-dimensional Schr\"odinger problem. Then, if we neglect vertex terms associated with the boundary conditions at the interface between open and closed portions of DNA, each integral is reduced to a  trace of the harmonic hamiltonian~\cite{Kleinert}. We obtain
\begin{eqnarray}
{\cal I}[x(s)]= \prod_{\{L^{o,c}\}}(\beta \lambda^{o,c}/a)^{-L/a}\left[\sum_{n=0}^{\infty}e^{-L^{o,c}\sqrt{f/\lambda^{o,c}}(n+1/2)}\right]^2 \nonumber \\
\label{factor2}
\end{eqnarray}
where  the factor $(\beta \lambda^{o,c}/a)^{-L/a}$  is a normalization term coming from the transformation of the path integral into a trace~\cite{Kleinert} and having the form of an equipartition, while the sums correspond to the aforementioned traces in each portion of open/closed DNA, and are simple geometric series. To sum them we take the approximation of strong tension, or $f \langle L^{o,c}\rangle^{2}/\lambda^{o,c}\gg1$, where $\lambda^{o,c}$ and $\langle L^{o,c}\rangle$ are, respectively, the bending rigidities and the average lengths for the open/closed portion considered. Then the expression in (\ref{factor2}) can be approximated as
\begin{eqnarray}
{\cal I}[x(s)]\simeq \prod_{i}\exp{\left\{-a\sqrt{f/\lambda(x_i)}-\ln \left[\beta \lambda(x_i)/a\right]\right\}},~~
\label{factor3}
\end{eqnarray}
 which concludes our integration. 
 
 It follows from (\ref{factor3})  that in the approximation of strong tension the effect of bending modes is to contribute the term 
%
$T\sqrt{f}\Delta (1/\sqrt{\lambda})-(T/a)\ln({\lambda^c}/{\lambda^o})$
%
to $W_{0,f,T}$ in (\ref{W0}), and a term 
%
$T\sqrt{f/{\lambda^o}}-({T}/{a})\ln(2\pi T a/\lambda^o)$
%
to $\Delta$ in (\ref{Delta}).

Let us keep track of the approximations leading to the expression (\ref{factor3}). First of all we project on the lowest eigenvalue, under the assumption that the force and the length of the open/closed portions are large enough. The condition  $f \langle L^{o,c}\rangle^{2}/\lambda^{o,c}\gg1$ simply implies that the tensile energy on open and closed portions of DNA exceeds its  elastic energy for a flexion of curvature $\langle L^{o,c}\rangle^{-1}$. This is consistent with treating flexural modes perturbatively---as we have assumed in their introduction in the previous section---with respect to a tension strong enough to prevent plectonemes.  

Equivalently the condition of strong tension can be seen as a lower limit for the minimal length of each open or closed portion of DNA: $\langle L\rangle^2\gg T l_p/f$, where $l_p=\lambda/T$ is the persistence length. For helical DNA, $l_p\simeq 50$ nm, and at room temperature ($T=4$ pN$\times$nm) one has, for a tension of 10 pN, $\langle L^c\rangle > 4$ nm, which seems quite reasonable. For open portions the persistence length is $\simeq$ .75 nm~\cite{Smith96}, and we find that the average length of bubbles must exceed a fraction of a nanometer, which is always satisfied because of the discrete nature of the problem. Larger tensions further improve the approximation.

 A further approximation has consisted in disregarding the integral over $\vec\psi$ at the boundaries of open and closed portions
 , which mathematically corresponds to neglecting the contribution of flexural modes to the ``vertex--factors'', or the free energy of the kink between open and close configurations. We are therefore assuming that most of the energetics of these kinks is controlled by the interplay between the stacking potential and the base bonds, with the only effect of bending being weakening the base bond itself.  
 
Finally, let us note that for dsDNA  (\ref{factor3}) becomes exact with $\lambda(x)=\lambda^c$, and the approximations are justified in the thermodynamic limit. One can then find the contribution of flexural modes to the stretch as $\Delta h=T\partial_f \ln {\cal I}$ as
\begin{equation}
\frac{\Delta h}{N a}=-\frac{T}{2\sqrt{f \lambda^c}}
\end{equation}
the same result obtained from the wormlike-chain model~\cite{Marko95, Odijk95} and supported experimentally~\cite{Wang97}. 

\section{Thermomechanical Stability}

We have so far introduced a mechanical model for open and closed sections of DNA and we have shown  that its thermomechanics reduces to the thermodynamics of a special PB or PBD model, whose parameters contain informations on the effects of fields, as well as on torsional and flexural rigidities for open and closed torsions. We now apply the formalism, produce predictions, and compare them with experimental results.

The strength of the potential $W_{\tau,f,T}$ controls the phase diagram for the stability of DNA. Indeed in general,  for a PB model such as the one in (\ref{Z}),  the condition for stability requires it to be larger than some monotonically increasing function of temperature~\cite{PB,PBD}, or
\begin{equation}
W_{\tau,f,T}\ge g(T),
\label{g}
\end{equation}
where the functional form of $g(T)$  depends on the particular potential chosen and can be computed numerically or analytically with certain approximations.  Here, in the context of the harmonic PD model, we  proceed to a continuum limit in which $i\to s$, $a\sum_i\to \int ds$,  and $\Delta x_i/a\to x'(s)$. Then, neglecting an irrelevant equipartition factor, $Z$ in (\ref{Z}) can be written as the propagator in imaginary time for an equivalent Schr\"odingier problem~\cite{Kleinert}, and thus as proportional to the trace,
\begin{equation}
Z \propto  e^{-\beta N a\Delta} ~\mathrm{Tr}~e^{-N a \hat H},
\label{Ztrace}
\end{equation}
  of the operator
\begin{equation}
\hat H=-\frac{1}{2k \beta}\frac{d^2}{dx^2}-\beta\chi(x)W_{\tau,f,T}.
\label{H}
\end{equation}
In the thermodynamic limit of large $N a$, (\ref{Ztrace}) projects on the lowest bound eigenvalue. Then, if the Schr\"odinger problem admits a purely continuum spectrum for certain values of the parameters, the disappearance of the last bound state corresponds to a critical surface for DNA stability~\cite{PB, Nisoli09}.

The problem in (\ref{H}) describes a quantum particle on a half-line with an attractive  square potential, which indeed admits a purely continuum spectrum. The problem is solved in standard textbooks~\cite{Landau} and one finds that DNA is stable for
\begin{equation}
W_{\tau,f,T} \ge \pi^2 T^2/8k{x^c}^2=T^2/\zeta
\label{g2}
\end{equation}
(here $\zeta=8k{x^c}^2/\pi^2$, has the dimension of a force per length square). Then one can  directly plot the phase diagram for stability, as in Fig.~\ref{Surface} and ~\ref {ft}, where parameters were fitted to  experimental data from~\cite{Bryant}, and which will be explained below. Before, we provide some heuristic considerations. A look at (\ref{W}) shows that, at constant tension and temperature, an unwinding torque ($\tau<0$) always destabilizes DNA by lowering $W$, whereas  a winding torque initially stabilizes it. Indeed a small positive torque stabilizes  DNA even at temperatures above unloaded denaturation (Fig~\ref{ft}), a property exploited by thermophile bacteria living at high temperatures~\cite{Watson}. As expected, negative torque destabilizes DNA, a mechanism exploited in biology for DNA opening and replication. 
	 
At fixed temperature, tension mainly destabilizes DNA via the term $-f v$ which describes the gain in tensile energy when bases are open (as previously described). Tension however also has an initially stabilizing effect by suppressing the thermal effect of flexural modes, described by the term $T\sqrt{f}\Delta \frac{1}{\sqrt{\lambda}}$). If positive torque is present, tension has another stabilizing effect, purely mechanical, by counterbalancing positive applied torque with a negative effective torque, given by (\ref{taueff}): this explains the skewness of the phase diagram toward positive torques. 

We now explain in more detail how the diagrams of Fig.~\ref{Surface} and~\ref {ft} are obtained.
By putting $f=0$, $\tau=0$, $T=T_D$ in (\ref{g2}), where $T_D$  is the denaturation temperature in the absence of torque or tension, we obtain the relationship 
\begin{equation}
T_D^2=\zeta\left(V_0-\frac{1}{2} \nu~\Omega^2-\frac{T_D}{2a}\ln\frac{\nu+ \mu}{\nu}-\frac{T_D}{a}\ln\frac{\lambda_c}{\lambda_o}\right)
\label{TD}
\end{equation}
 that relates $V_0$, $k$, and $x^c$ via  $\zeta=8k{x^c}^2/\pi^2$. These quantities are hard to relate to measurable physical parameters, due to the simplicity of the model. While $V_0 a$ represents the average energy of the base-bond, its value is renormalized by angular and flexural fluctuations, and can hardly be disentangled by the effect of the stacking potential in any realistic treatment. Also,  $k$ and $x_c$ are poorly defined: indeed the complexity of the stacking potential goes well beyond the purely elastic term we have introduced, whereas $x_c$, the length above which the bond breaks cannot be measured precisely. Therefore (\ref{TD})  serves the purpose of eliminating $V_0$ from our equations, since $T_D$ is generally known (we will take $T_D=350$ K~\cite{Saenger}). Then, we will see, only $\zeta$ needs to be fitted, and its value controls the sharpness of the transition.

From (\ref{g2}) and (\ref{TD}) we can eliminate $V_0$ and readily obtain the equation for the critical surface:
\begin{eqnarray}
&&\tilde\Omega \tau -\frac{1}{2} \frac{\tilde \mu 
}{(\tilde \nu+\tilde \mu)\tilde \nu} \tau^2-v f +\frac{\tilde\nu+\tilde \mu}{2}(\Omega-\tilde\Omega)^2 +\nonumber \\
&&+T\sqrt{f}\Delta \frac{1}{\sqrt{\lambda}}+\frac{T_D^2-T^2}{\zeta}+\frac{1}{2a}\left(T_D l -T \tilde l\right)=0,
\label{r}
\end{eqnarray}
where  $l=\ln[(\nu+\mu)/\nu]+2 \ln[{\lambda^c}/{\lambda^o}]$ and $\tilde l=\ln[(\tilde\nu+\tilde\mu)/\tilde\nu]+2\ln[{\lambda^c}/{\lambda^o}]$  are dimensionless. We fit our parameters to the experimental data from Ref~\cite{Bryant}. There are only four parameters needing to be fit: a typical value for the denaturation temperature used in theoretical treatments~\cite{Cocco, Barbi} is $T_D=350$ K~\cite{Saenger};   data on torque-winding experiments from Ref~\cite{Bryant} give us $\mu=1.4 \times 10^3$ pN; $\lambda^c$ can be obtained from the persistence length of DNA, which is well known from literature to be around $50$ nm. There is no such agreement for the persistence length of ss-DNA, but most data give it as less than 1 nm~\cite{Smith96}, and we fit it to $0.5$ nm. As already explained  above, for the geometrical parameters we take $R\simeq10$ \AA~for the radius of the DNA molecule,  $h_0\simeq3.4$ \AA~for the elevation between consecutive nucleotides, $a\simeq 7$ \AA~ for their distance along the backbone, and $\Omega=2\pi/10$ for the rotation per base pair of DNA (Fig. 1). This leaves $n$, $\zeta$ and $\nu$ to be fitted. The first depends on the effective radius $r$ of twisting for the open backbones, as described above. The second and the third are hard to relate to physical observables (also above). Choosing   $\zeta=1.3 \times10^3$ pN$\times$nm$^2$, $\nu=26$ pN and $n=0.23$ (which implies $r=3.5$ \AA, a reasonable value), provides a remarkably good fit for 10 experimental data points (20 numbers) from Ref.~\cite{Bryant} in the $f$ {\it vs.} $\Gamma$ phase diagram of Fig.~\ref{ft}, by effectively fitting only 4 numbers:  $\lambda^c$, $\nu$, $n$, $\zeta$.

\begin{figure}[t!]   
\center
\includegraphics[width=3 in]{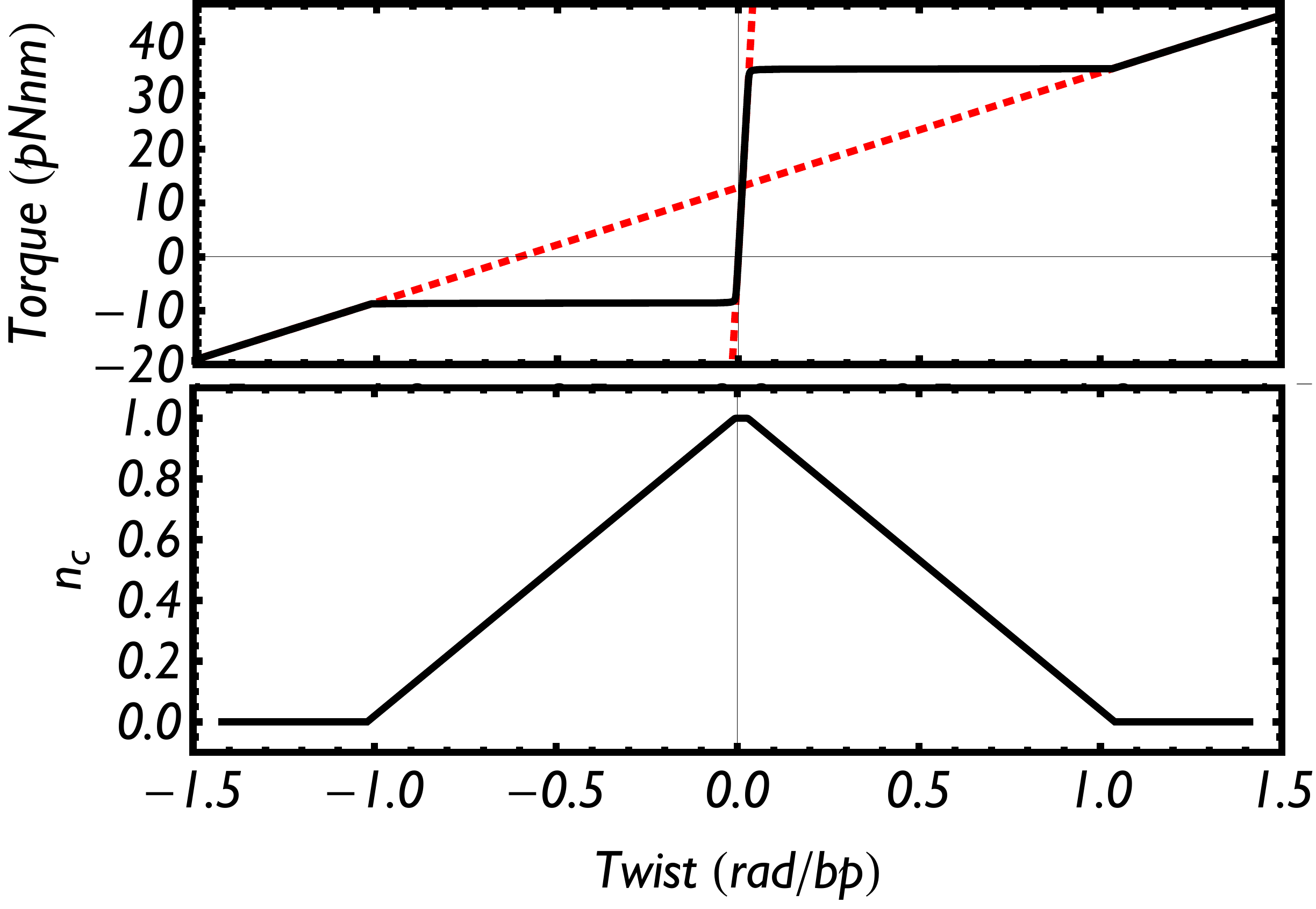}
\caption{Top: Plot of predicted curve of applied torque vs. twist of a DNA strand beyond denaturation; the red lines represent the purely mechanical twist of (\ref{Deltathetac0}). Bottom: the density of closed base pairs $n^c$  (or fraction of dsDNA) plotted vs. twist, as computed from (\ref{ncaverage}): during transition, corresponding to the critical value of the torque at which the twist jumps, the density of closed base pairs is linear in the twist, suggesting a first order phase transition.}
\label{totw2}
\end{figure}

Because  $W_{\tau,f,T}$ in  (\ref{W}) is quadratic in $\tau=\Gamma/a$, there are  two critical torques for denaturation, a negative ${\Gamma^c}^-$ and a positive ${\Gamma^c}^+$, which naturally depend on tension and temperature, as seen experimentally~\cite{Bryant}. From (\ref{W}) and also (\ref{r}) we see that there is  an optimal positive torque, independent of temperature, that maximizes $W_{\tau,f,T}$, and it is given by
\begin{eqnarray}
\Gamma_m= a \frac{\tilde\mu+\tilde\nu }{\tilde\mu}\tilde \nu ~\tilde \Omega\simeq a\nu ~\Omega +f\left(n\Omega-\frac{o\nu}{\mu
}\right).
\label{Gm}
\end{eqnarray}
Then from an experimental point of view,  $\Gamma_m$ can be easily computed  as the middle point between critical torques at  given tension, or
\begin{equation}
\Gamma_m=({\Gamma^c}^+ + {\Gamma^c}^-)/2.
\end{equation}
 It follows from (\ref{Gm}) that the critical lines at fixed temperature in the tension vs. torque space posses a skewness for positive torques, seen in Fig.~\ref{ft}, which corresponds to experimental observation~\cite{Bryant}. From (\ref{Gm}) we see that this skewness results from backbones twisting under torque in portions of open strands: indeed it would be erroneously negative for $r=0$ and therefore $n=0$. From experimental data in Ref~\cite{Bryant} that place torque-induced transitions at ${\Gamma^c}^+=34$ pN$\times$nm, ${\Gamma^c}^-=-10$ pN$\times$nm for a tension $f=15$ pN, and at $f=60$ pN,  for ${\Gamma^c}^+=33$ pN$\times$nm~\cite{Bryant}, we can predict for that experiment a melting at zero torque and tension $f\simeq60$ pN (blue square in Fig.~\ref{ft}) in good agreement with the experimentally observed  overstretching transition. We will see in the next sub-section that our analysis implies a force-induced melting~\cite{Mameren} rather than a transition to a double helix with distortions~\cite{Ha}.

In Fig.~\ref{Tf} we  use (\ref{r}) to predict critical temperatures as a function of the applied loads. From (\ref{r}) we see that  $T^c$ is maximized for $\Gamma=\Gamma_m(f)$ given by (\ref{Gm}), which therefore corresponds to {\it the most stable configuration at any temperature}, for a given tension $f$, as shown in the plots of Fig.~\ref{Tf}.  We observe that---as expected---a low tension stabilizes DNA, by suppressing flexural modes:  its contribution $T\sqrt{f}\Delta\frac{1}{\sqrt \lambda}$ being positive and scaling as a square root. However, at larger tensions DNA is destabilized by the $-fv$ contribution to $W_{\tau,f,t}$:  this term is related to the unwinding induced by tension, and the gain in tensile energy when portions of DNA open, since $v=1-h_0/a$ is the ratio between lengths of open and closed DNA. Tension can stabilize DNA in a purely mechanical way in the regime of large positive applied torque, which it counterbalances, as in (\ref{taueff}). Finally, Fig.~\ref{Tf} shows that rather high critical temperatures can be achieved under appropriate mechanical load. While this effect could be an artifice of the approximations employed in summing the flexural modes in section 3.2, it is nevertheless true that hyperextremophiles have been found to survive at high temperatures. While it has been speculated that their stability is genomic in origin, recent findings appear to challenge the hypothesis~\cite{Hurst}. In particular, ``strain 121'' (Geogemma barossii) has been found to live, albeit biostatically, at temperatures above $130~^o$C~\cite{Kashefi}.

\section{Non-Linear Mechanochemical Response}

Having addressed the problem of DNA stability, we now consider the question of response to fields.  Experiments on DNA single molecule manipulations allow us to access  measures of stretch vs. tension or supercoiling vs. torque curve. These same quantities can be computed in the context of our model. In the following we will show that  thermomechanical responses can be obtained from the purely mechanical ones via knowledge of the density of open base pairs.

From $\tilde \omega=\Delta \theta-\tilde \Omega$ and $\langle \tilde \omega \rangle={(N a)}^{-1}T \partial_{\tau} \ln Z$, we have
\begin{equation}
\langle \Delta \theta \rangle=\tilde\Omega+(Na)^{-1} T \partial_{\tau} \ln Z.
\label{omegaav}
\end{equation}
Similarly for the average elongation
\begin{equation}
\langle \Delta h \rangle= T \partial_{f} \ln Z.
\label{stretchav}
\end{equation}
One sees immediately the effect of the term $\Delta$ in (\ref{Z}) and (\ref{Ztrace}) to the average supercoiling and elongations: in regions of the phase space in which DNA denatures, there are no bound states and therefore the only extensive part of the partition function in (\ref{Ztrace}) is $Z \propto  e^{-\beta {N a}\Delta}$. From (\ref{omegaav}) we have then $\langle \Delta \theta\rangle=\tau/\tilde \nu$, the same formula obtained before for open portions of DNA in (\ref{Deltathetac0}).

To compute supercoiling of DNA below transition we need a knowledge on the density of closed bases.
Consider  the functional
\begin{equation}
{\cal N}^c[\{x_s\}]=\sum_{s=1}^N\chi(x_s),
\label{chi}
\end{equation}
which for every configuration of base distances $\{x_s\}$ returns the corresponding  number of closed bases. Then, from  (\ref{Z}), we find that  the average number of closed bases $n^c$ is a function only of $W$,
\begin{equation}
 n^c(W) = N^{-1} \langle {\cal N}^c\rangle=T (Na)^{-1}\partial_W \ln Z(\Delta,W),
\label{ncaverage}
\end{equation}
and clearly $n^o=1-n^c$ is  the average number of open bases; both quantities are computed in the context of the PB model and only depend of $W$, the depth of the square potential. As $W$ depends on loads and temperature, so do $n^c$, $n^o$. 
Equation (\ref{ncaverage})  allows us to relate the thermomechanical responses to the purely mechanical one. From  (\ref{omegaav}) and  (\ref{ncaverage}) we find  
\begin{equation}
\langle \Delta \theta \rangle= \tilde \Omega -\partial_{\tau} \Delta +n^c \partial_{\tau}W_{\tau,f,T},
\label{omegaav2}
\end{equation}
which, using $n^o+n^c=1$ and the expressions for $\Delta$ in  (\ref{Delta}) and $W_{\tau,f,t}$ in (\ref{W}) returns
\begin{equation}
\langle \Delta \theta \rangle= n^o \frac{\tau}{\tilde \nu}+n^c\left( \tilde \Omega + \frac{\tau}{\tilde \nu+\tilde \mu}\right).
\label{omegaav3}
\end{equation}
We recognize in (\ref{omegaav3}) the expression for the purely mechanical angular responses relative to open and closed portions of DNA, obtained in (\ref{Deltathetac0}). The average angular displacement per base can thus be written in the following  intuitive way
\begin{equation}
\langle \Delta \theta \rangle=n^o \Delta \theta^o+n^c \Delta \theta^c.
\label{omegaav4}
\end{equation}
Then from (\ref{omegaav2}) one can obtain the supercoiling by solving for the PD model. The fact that the  expressions for $\Delta \theta^{o,c}$   are purely mechanical is a consequence of having truncated the hamiltonian at the quadratic order  in the angular fluctuations: then the thermal average of the angular displacement in closed or open portions is simply the minimum of the hamiltonian, which of course corresponds to the purely mechanical response. Then the overall average $\langle \Delta h \rangle$ is simply given by the contribution of open and closed portions, whose relative density is controlled by the PB model-- and this is what is stated by (\ref{omegaav4}) and (\ref{stretchav}). 

\begin{figure*}[t!]   
\center
\includegraphics[width=5.5 in]{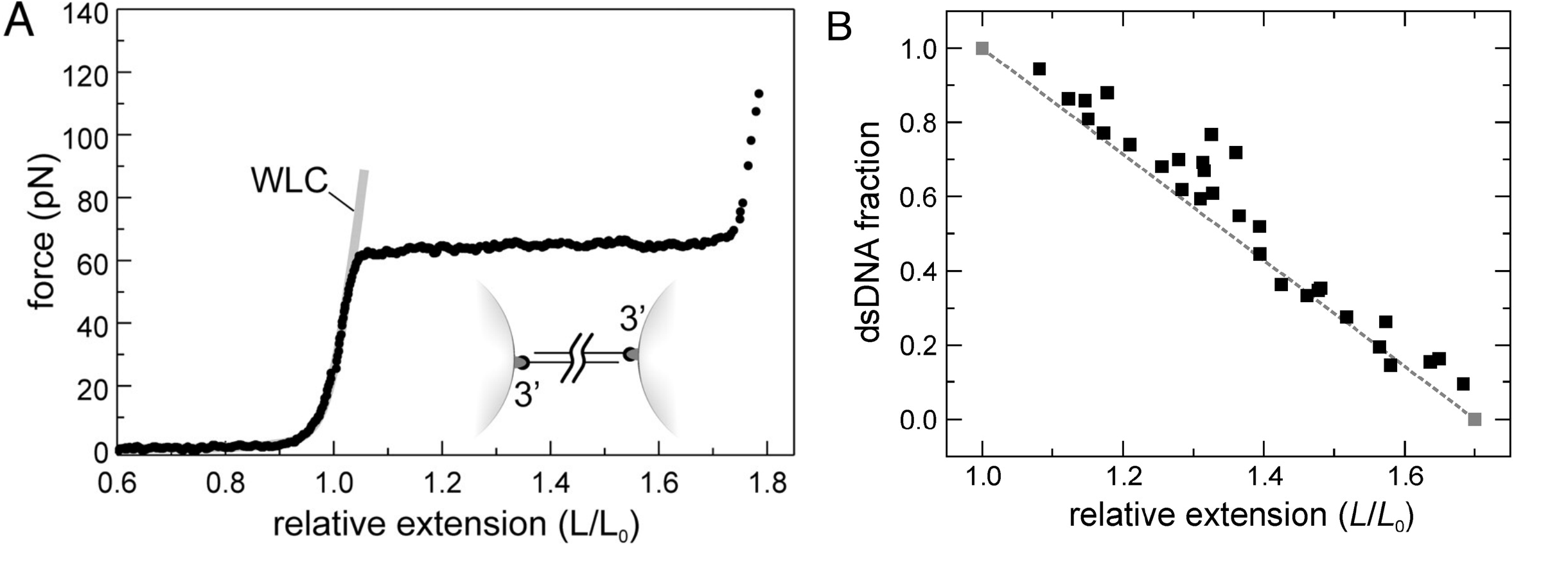}
\caption{Experimental results on overstretching transition reprinted with permission from Ref~\cite{Mameren}. (A) Typical force-extension curve of a 3'-3' attached DNA, with free 5' ends (schematically represented in the inset). At 65 pN, the DNA molecule undergoes the  transition, corresponding to an 170\% increase in length. (B) The fraction of dsDNA, obtained from the length of YOYO-labeled segments, plotted as a function of DNA extension.  The gray dashed line connecting the two gray points indicates a linear behavior in the stretch, completely analogous to the one reported in Fig.~\ref{totw2} for torque-induced melting.(WLC stands for wormlike chain model.)
}
\label{Mameren1}
\end{figure*}

The situation for the stretch is somewhat different. Clearly we can follow an analogous treatment and write 
\begin{equation}
\langle \Delta h \rangle=n^o \Delta h^o+n^c \Delta h^c,
\label{stretchav}
\end{equation}
where, from (\ref{stretchav}) and (\ref{ncaverage}), the average stretch for open and closed portions are given by
\begin{eqnarray}
\frac{\Delta h^o}{{N a}}&=&-\partial_f \Delta \nonumber \\
\frac{\Delta h^c}{{N a}}&=&-\partial_f( -\Delta+W).
\label{stretchoc}
\end{eqnarray}
However,  (\ref{stretchoc}) are not purely mechanical as they contain temperature-dependent terms from the contribution to elongation by flexural modes, which, as described above, correspond to a wormlike chain model. Indeed the reader can verify that this  corresponds to an {\it extensible} wormlike chain model, as both the closed and open portions of DNA can elongate by altering their winding angle.

Following the continuum approximation, we obtain $n^c$ from the lowest bound eigenvalue $\kappa(W,T)$ (which has the dimension of a reciprocal length) of the operator in (\ref{H}). From (\ref{Ztrace}) we have
\begin{equation}
n^c(\tau,f,T)=-T\partial_W\kappa(W,T)|_{W=W_{\tau,f,T}},
\label{nc2}
\end{equation}
and $\kappa(W,T)$ can be solved for exactly:  the algebra that leads to $\kappa$  can be found on standard textbooks~\cite{Landau}. However an  interesting point is worth discussing.

It generally assumed that  that the denaturation transition of DNA is first order (see Ref~\cite{Kafri} and references therein). The  PB model predicts a  smoother transition. The following  PBD model showed that the transition could be made sharp  via the introduction of a nonlinear stacking potential, which took into account that, when base pairs are open, no stacking interaction exists. While the same non-linear potential can--and should--be applied to our model (we will report on it elsewhere), even within a linear stacking potential, our model can produce transitions of desired sharpness.

While the transition predicted within our model with linear stacking potential is second order (the order parameter, the density of closed bases is continuous) it can closely mimic a first order transition.  This is because in our model the depth of the potential in (\ref{W}), unlike in the PB and PBD models, depends on temperature. The transition can then be made sharper by increasing the value of the parameter $\zeta$, which depends on $k$. Intuitively, raising $k$ means increasing the ``mass'' of the quantum particle described by (\ref{H}): in the classical limit, the particle is either  inside the step-like potential, or outside it, and the transition is first order. A look at (\ref{TD}) and (\ref{r}) shows that  both  denaturation transition and  load-induced transition can happen in the limit of $\zeta \to \infty$. As discussed in the previous sub-section, the fitting of data for stability of DNA lead to a rather large value   $\zeta=1.3 \times10^3$ pN$\times$nm$^2$, which then leads to sharp transitions. 

\begin{figure*}[t!]   
\center
\includegraphics[width=5.5 in]{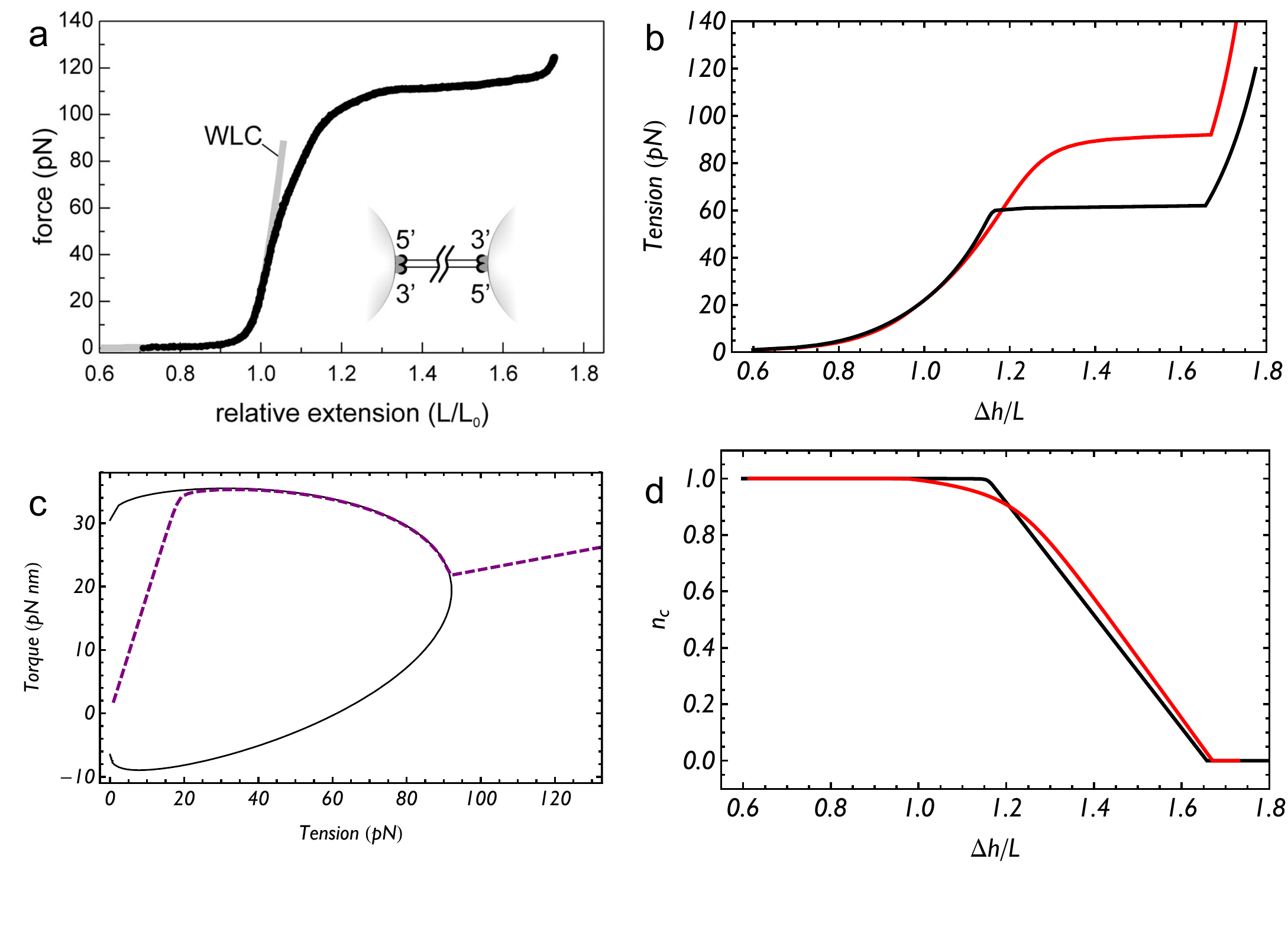}
\caption{(a) Experimental results on overstretching transition reprinted with permission from Ref~\cite{Mameren}:  In a 3'5'-5'3' attachment geometry, where all four strand ends are linked and DNA cannot rotate, the transition is smoother then overstretching with free ends (see Fig.~\ref{Mameren1}, and the critical force increases considerably, to 110 pN. (b) Our predictions for the overstretching transition with constrained (black line) and unconstrained (red line) ends, with our model parametrized on a different experiment~\cite{Bryant} qualitatively agree with Ref.~\cite{Mameren}: the constrained transition is smoother and happens at higher critical tension. (c) The
curve $\Gamma_{\mathrm{cons}}(f)$ (purple, dashed) describing the torque exerted by the apparatus on the DNA with fix ends follows the critical line (black, solid), thus explaining the initial smoothness of the transition. (d) Fraction of dsDNA vs. stretch for the constrained (black line) and unconstrained (red line) DNA. (WLC stands for wormlike chain model.)}
\label{Mameren2}
\end{figure*}

\subsection{Torque-induced melting}

We can thus obtain, analytically, the angular response of DNA to applied fields. In the experiments of torque-induced melting, the filament is held at a given tension (to avoid plectonemes)  and supercoiled while the applied torque and the resulting average deviations are recorded.  The nature of the transition can be seen in Fig.~\ref{pappa}, where we plot  experimental results from Ref.~\cite{Bryant} for  torque vs. twist, along with our predictions. The transition observed experimentally is  sharp, and so are our findings.

Our framework allows to also compute the fraction of dsDNA at given tension, torque and temperature.  A proof of the first order nature of the transition is given by Fig.~\ref{totw2}, in which we plot torque vs. twisting until beyond denaturation and compare it with the fraction of dsDNA: the density of close bases decreases linearly with the increase in over- or under-twist. Indeed   (\ref{omegaav4}) shows that, if the transition is of first order, then at the critical value of the external loads the mechanical twist $\Delta \theta^{o,c}$ does not change, and $\langle \Delta \theta\rangle$ is linear in $n^c$.

\subsection{Overstretching transition}

A result analogous to the one presented in Fig.~\ref{totw2} has been recently reported by Mameren and collaborators~\cite{Mameren}, in the context of the over stretching transition (Fig~\ref{Mameren1}). The experiment was performed to shed light on the lengthy discussion concerning the nature of the overstretching transition~\cite{Romano}, which has generated two qualitatively different models: in one the transition involves a structural conversion to bound, double stranded conformation, or S-DNA~\cite{Cluzel}, consisting of partially (or fully) unwound   DNA but with still base pairing; in the other, DNA simply melts, as in the thermal denaturation transition, and separates into to ssDNA~\cite{Wenner}.

By using YOYO, a dsDNA-specific fluorescent dye~\cite{Rye}, Mameren and collaborators were able to map the density of closed bases during the transition (Fig.~\ref{Mameren1}). They found a linear dependence between dsDNA fraction and  stretch, which  points quite unambiguously to a first order melting transition. The theoretical predictions for the overstretching transition based on our model are in line with their findings, as shown in  Fig.~\ref{Mameren2}: we too find a sharp transition at around $60$~pN, and a linear behavior in the density of closed pairs vs. stretch (Fig.~\ref{Mameren2}:d). 

Mameren et al. also found that the transition is nucleation limited. Very few early precursors are formed, coalescing at the free ends (or around nicks) and then propagating. As a further test  they performed stretch at fixed ends (Fig.~\ref{Mameren2}), where such coalescence cannot occur, and indeed found a much higher critical tension for transition ($\simeq110$ pN), which they ascribed to the removal to favorable nucleation points at the ends. 

However the overstretching transition for the 3'5'-5'3' attached DNA is less sharp than the one at 65 pN. Is that a ``more continuous'' kind of transition? Our treatment allows us to address this question. It is important to understand that  overstretching with unconstrained ends corresponds to zero applied torque. However, when the ends of the molecule are constrained such that no rotation is involved, the apparatus exerts a {\it tension-dependent} positive torque $\Gamma_{\mathrm{cons}}(f)$ on the molecule. Such constrained torque is obtained implicitly by setting $\langle \Delta\theta\rangle= \Omega$ in (\ref{omegaav4}). 

\begin{figure*}[ht!]   
\center
\includegraphics[width=2.5 in]{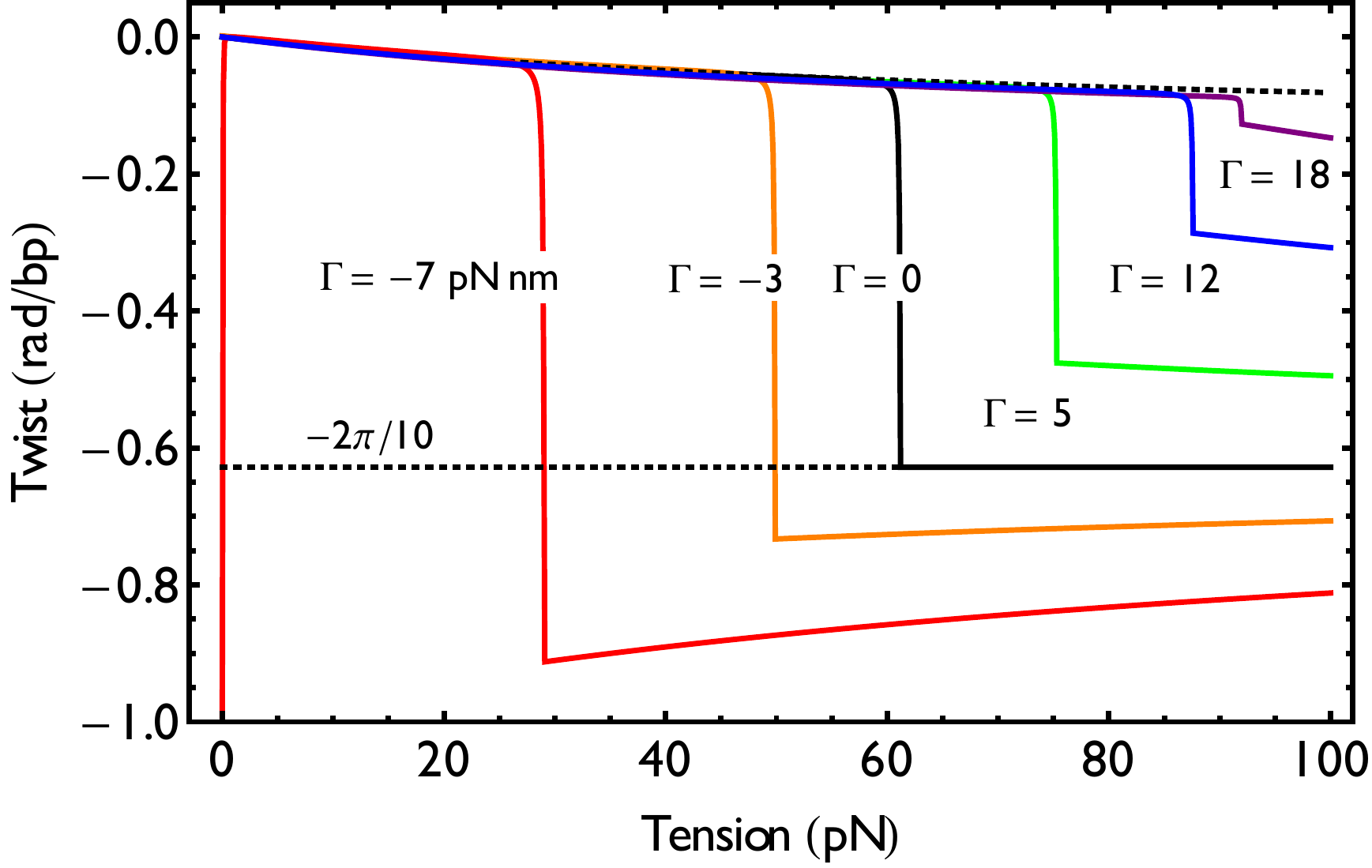}\hspace{5mm}\includegraphics[width=2.6 in]{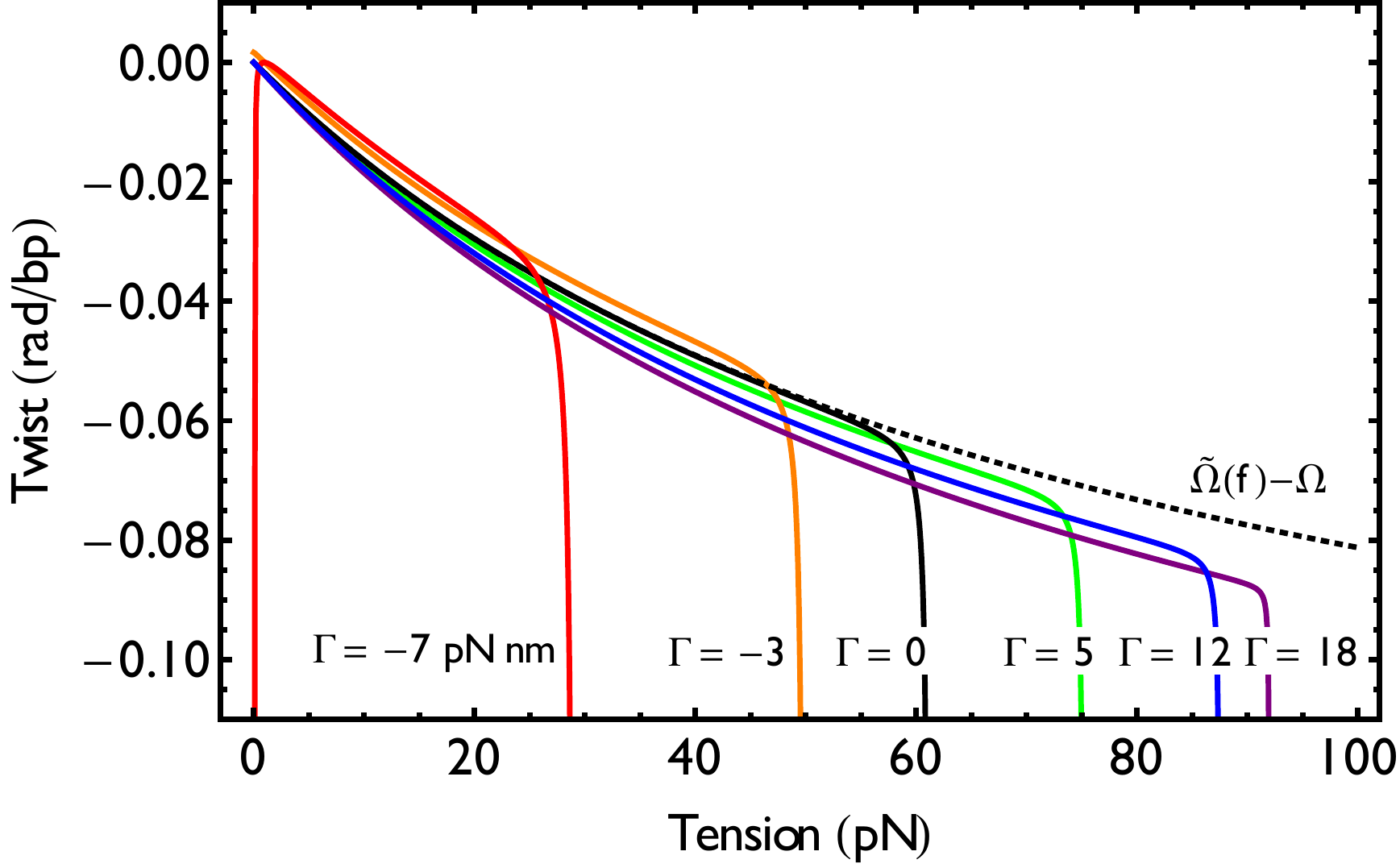}\vspace{5 mm}
\includegraphics[width=2.5 in]{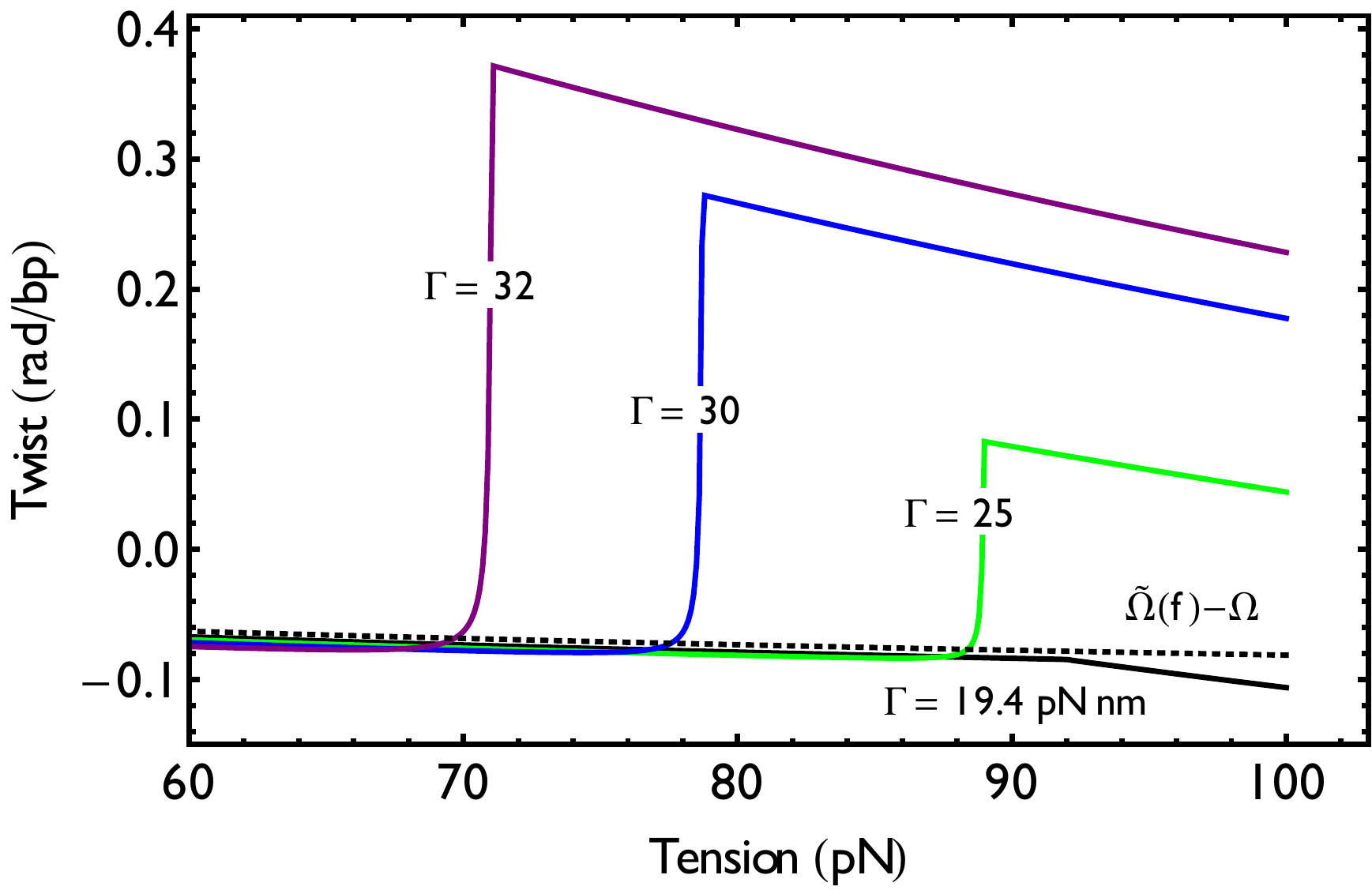}\hspace{5mm}\includegraphics[width=2.6 in]{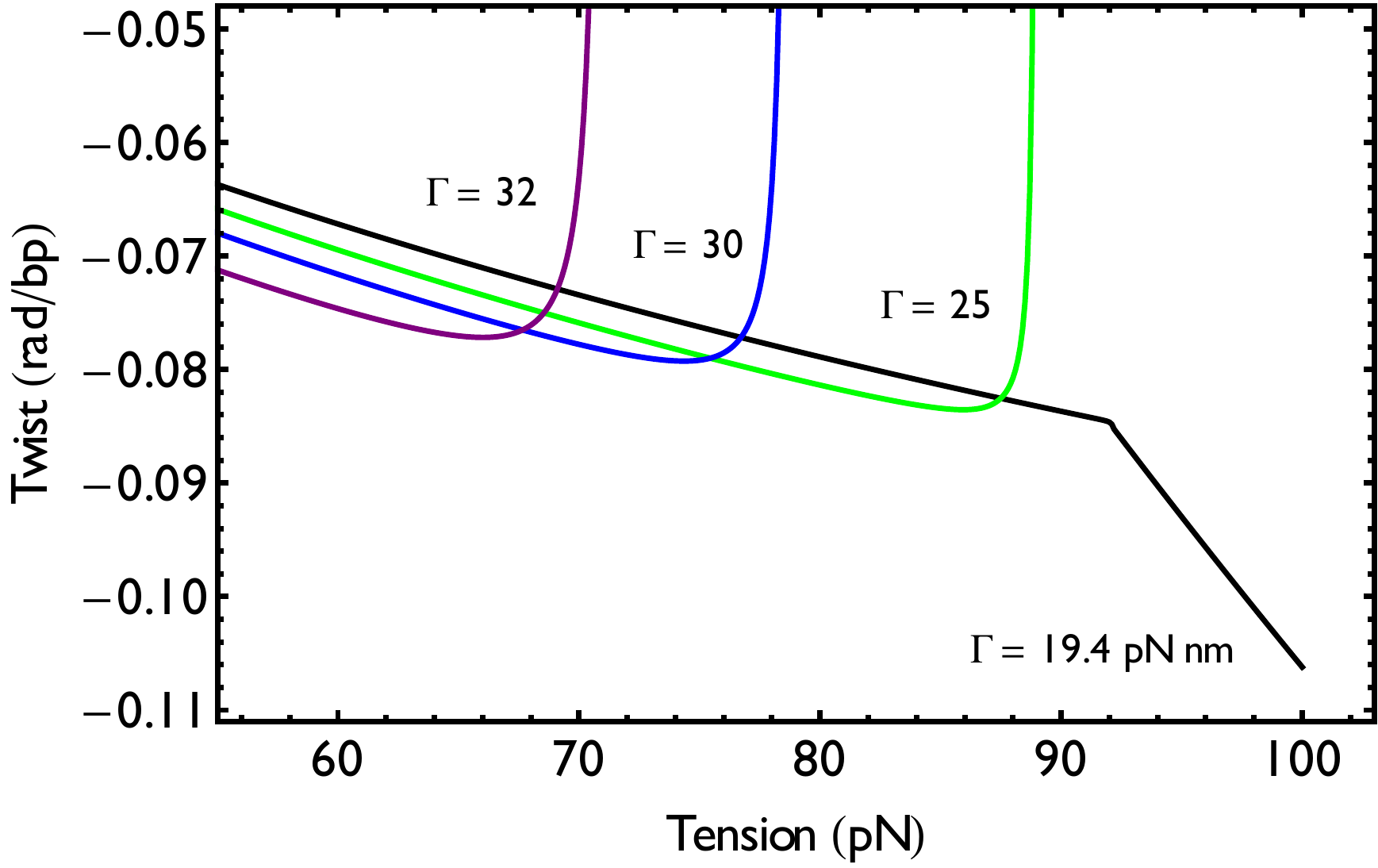}

\caption{Predictions for DNA twist as a function of applied tension at different torques, for torques  below (top two panels) and  above (bottom two panels) the ssDNA compensating torque, which with  our parameters corresponds to $\simeq19.4$ pN$\times$nm. The curves are shifted such that the twist at small tension is taken to be zero. Top Left: profiles of supercoiling  vs. tension  at torques below the ssDNA compensating value; the line corresponding to the overstretching transition (solid black) corresponds to zero applied torque, and shows a jump in supercoiling due to DNA melting;  above transition, the supercoiling increases (decreases) with tension for positive (negative) torque, due to the compensating effect of tension over torque acting on twisted ssDNA (Fig. 1). Top Right:  profiles of supercoiling  vs. tension  at torques at and above the ssDNA compensating value; the solid black line corresponds to traction under the ssDNA compensating torque, and correctly shows continuity with no jump in the supercoiling; at larger torques, the supercoiling jumps to positive value as tension cannot compensate torque in unwinding the twisted ssDNA strands. 
 Bottom Right: same curves plotted on a smaller scale, to reveal the behavior in the stable phase.}
\label{twt}
\end{figure*}

In Fig.~\ref{Mameren2}:c we plot the line described by $\Gamma_{\mathrm{cons}}(f)$,  the positive torque exerted by the apparatus on the constrained over stretching experiment, at room temperature. Together, we plot the phase diagram. Note that the torque grows linearly at the beginning, until it reaches the proximity of the critical line. Then it closely tracks the critical line until the end-constrained overstretching  transition takes place. After that, the curve returns to linear, but with a different slope. 

We can understand this behavior. Well inside the stability region DNA is mostly ssDNA, or $n^c=1$: therefore, in order to enforce the constraint $\langle \Delta\theta\rangle= \Omega$, the torque exerted by the apparatus,  $\Gamma_{\mathrm{cons}}(f)$,  must  compensate the effective torque exerted by the tension, given by $\tau_{\mathrm{eff}}$ in (\ref{taueff}). It follows that $\tau$ and thus $\Gamma_{\mathrm{cons}}$ must grow linearly with $f$, at least initially, following what we call the line of dsDNA compensating  torque.  This line would intersect the proximity of the critical line for melting at a tension of about 20 pN. Yet DNA does not melt there. That can also be understood by considering the expected behavior of the ssDNA, above transition. From (\ref{Deltathetac0}) the  ssDNA must obey $ \Gamma_{\mathrm{cons}}=a \Omega(\nu +n f)$, which corresponds to another, different straight line outside the  region of stability, which we call the line of ssDNA compensating  torque. The intersection of this second straight line with  the critical line clearly gives the point of transition.

But what happens when $\Gamma_{\mathrm{cons}}$ approaches the critical line from below the  transition, along the line of dsDNA compensating torque? As $\Gamma_{\mathrm{cons}}(f)$ reaches the proximity of the critical line, bases start to open and ssDNA with lower torsional rigidity appears. These open portions require less compensating torque from the apparatus, as their torsional rigidity is one order of magnitude lower, thus reducing the torque exerted by the apparatus. The changing admixture of ssDNA and dsDNA then allows for the curve  $\Gamma_{\mathrm{cons}}(f)$ to follow the critical line, until it reaches the critical point described above, and can denature. Therefore, the fact that the molecule resides in the proximity of the transition for a large interval of tension explains the smoothness of the transition at fixed ends. In reality the transition takes place under a calibrated tension-dependent torque generated by the constraint.

\begin{figure*}[t!]   
\center
\includegraphics[width=2.5 in]{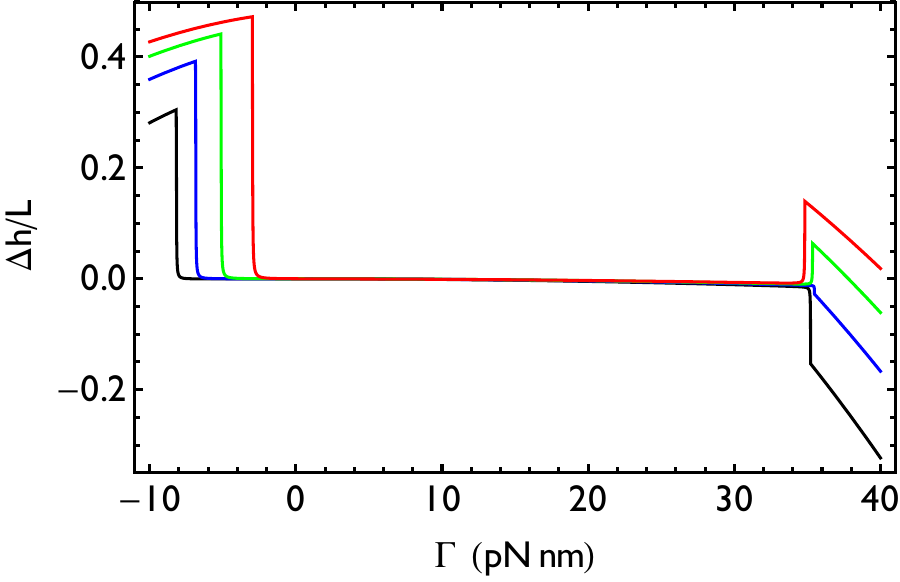}\hspace{5mm}\includegraphics[width=2.6 in]{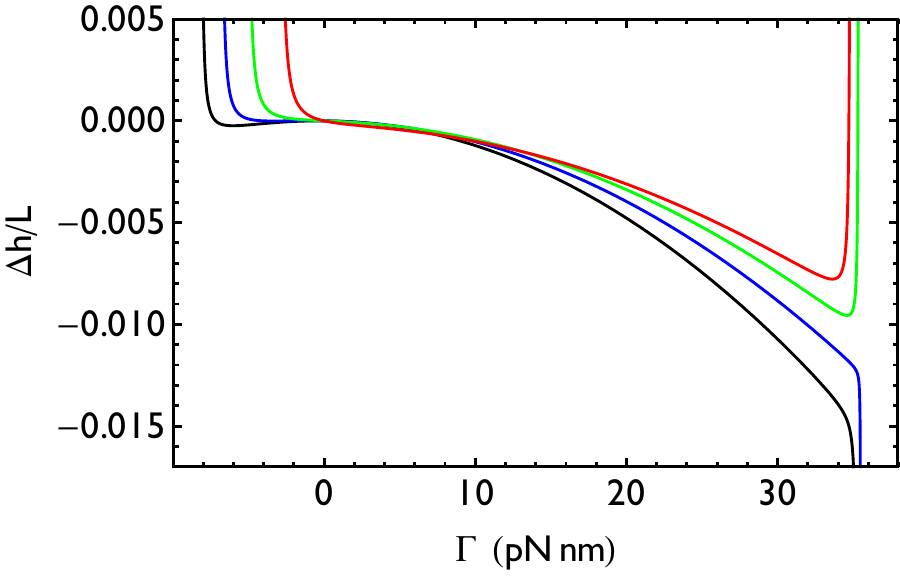}\vspace{5 mm}
\includegraphics[width=2.5 in]{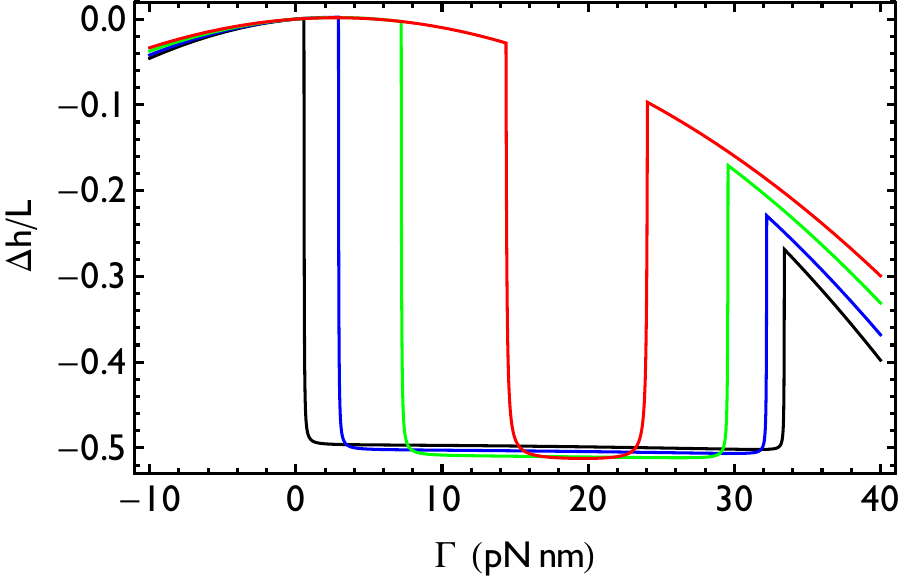}\hspace{5mm}\includegraphics[width=2.7 in]{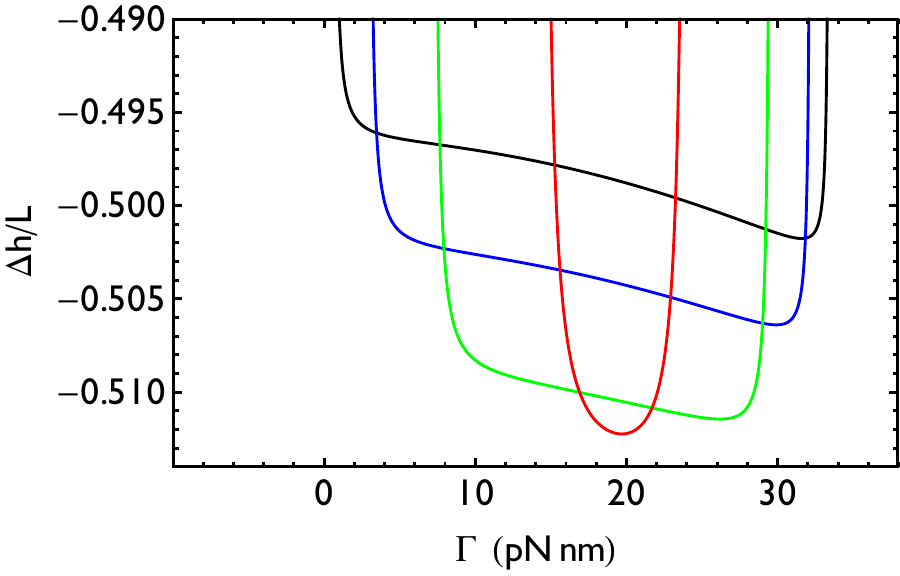}

\caption{Predictions for stretch as a function of applied torque at different tension, for tensions  below (top two panels) and  above (bottom two panels) the overstretching value. The curves are shifted such that the stretch at zero torque is taken to be zero. Top Left: stretch profiles vs. torque at tension  of 20 (black), 30 (blue), 40 (green), 50 pN (red); note that at the overwinding transition the stretch jumps toward elongation if the tension is large enough to compensate the shortening effect of torque on the ssDNA, otherwise the molecule shortens. Top Right: same curves plotted on a smaller scale, to reveal the behavior in the stable phase. Bottom Left: stretch profiles vs. torque at tension above the over stretching  tension of 60 (black), 70 (blue), 80 (green), 90 pN (red); at these tensions  DNA is melted unless torque is applied; under torque it goes back to ssDNA and thus contracts; for  torques larger than critical it stretches again.  Bottom Right: same curves plotted on a smaller scale, to reveal the behavior in the stable phase.}
\label{dhg}
\end{figure*}

\subsection{Profiles of mechanochemical response}

In the previous two subsections we  have dealt with known and celebrated cases. We conclude this section by providing further general predictions.

In Figure~\ref{twt} we provide predictions for DNA twist as a function of applied tension at different torques, for torques  below (top two panels) and  above (bottom two panels) the ssDNA compensating torque. This latter is given by the intersection between the critical line and the line of ssDNA compensating torque, described in the previous subsection, and with  our parameters corresponds to $\simeq19.4$ pN$\times$nm. The line of overstretching transition (solid black) corresponds to zero applied torque, and shows a jump in supercoiling due to DNA melting, but no further change in twist after melting, as no torque is applied. As we know already from the phase diagram, a negatively applied torque reduces the transition tension, whereas any applied $\Gamma$ such that $0<\Gamma<\Gamma_m$ given by $\ref{Gm}$ stabilizes DNA and increases the critical tension. Above transition, the supercoiling increases (decreases) with tension for positive (negative) torque, due to the compensating effect of tension over torque acting on twisted ssDNA (Fig. 1): torque will twist the two separated strands, tension will untwist it.  

\begin{figure*}[ht!]   
\center
\includegraphics[width=3 in]{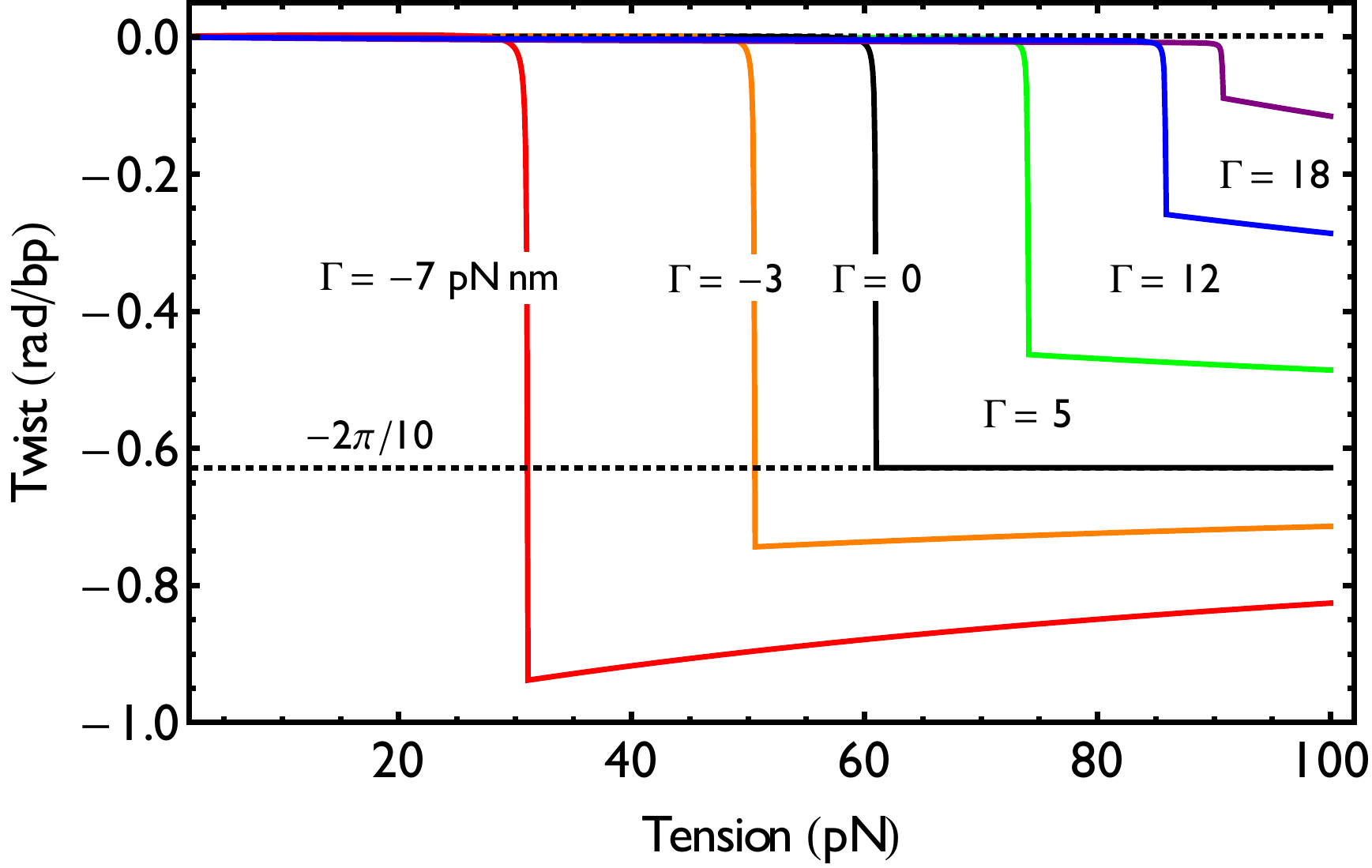}\hspace{5mm}\includegraphics[width=3.1 in]{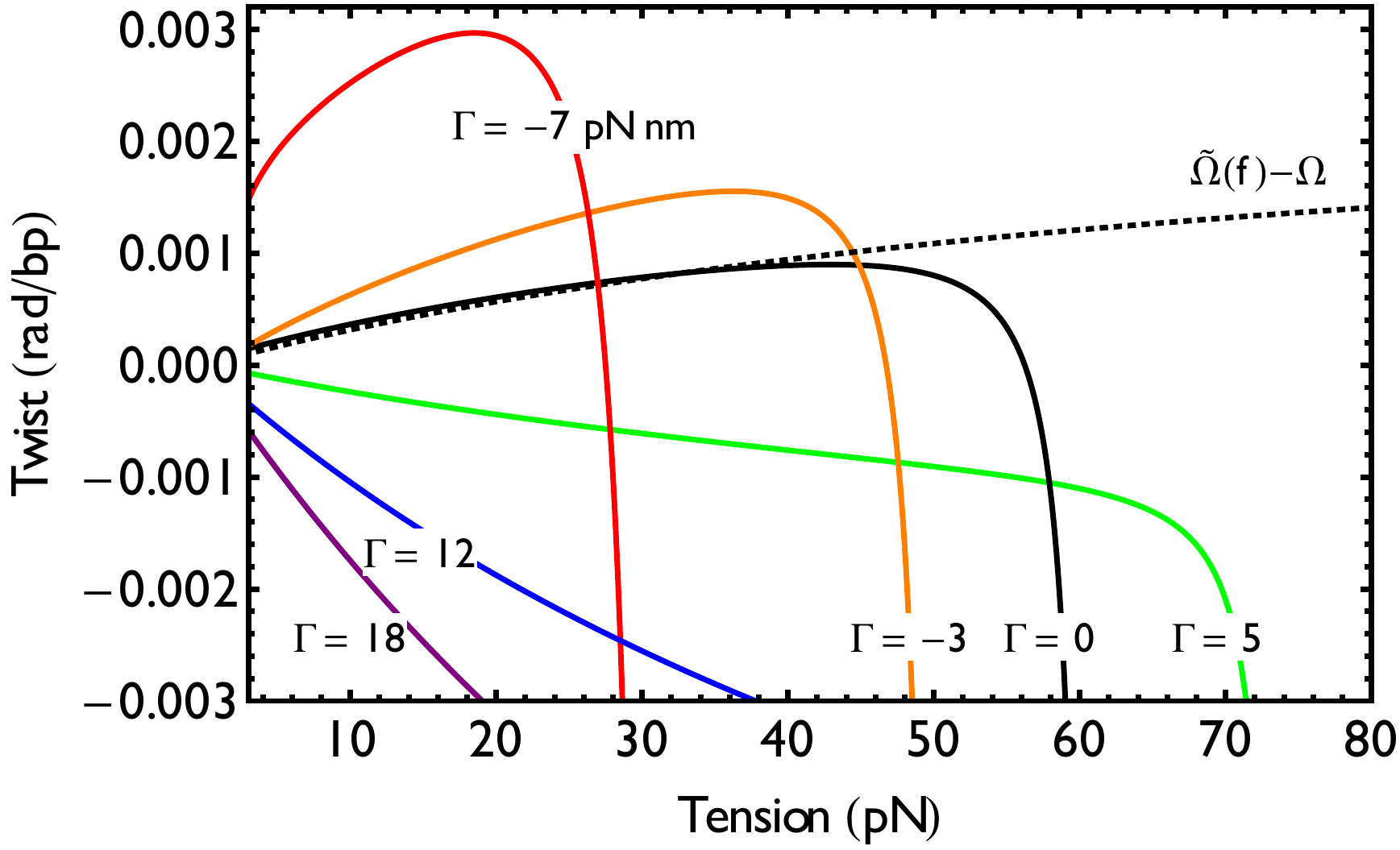}\vspace{5 mm}

\caption{Predictions for DNA twist as a function of applied tension at different torques, when the anomalous twist/stretch coupling~\cite{Bust2} is taken into account.  Left: profiles of supercoiling  vs. tension. Left: Same plots on a smaller scale reveal the overwinding under tension.}
\label{twto}
\end{figure*}

At  torques larger than $19.4$ pN$\times$nm, the supercoiling jumps to positive values, as tension cannot compensate torque in unwinding the twisted ssDNA strands. It is interesting then to observe the case of $\Gamma=19.4$ pN$\times$nm, which corresponds to traction under the ssDNA compensating torque. The plots correctly shows continuity  with no jump in twist: the applied torque is exactly right to compensate, in the ssDNA phase, the jump in twist at melting. 

Figure~\ref{dhg} reports stretch as a function of applied torque for different tensions at room temperature. In the region of DNA stability the change in elongation due to the applied torque is of the order of 1.5\% or less. However, as expected, the transition to melted DNA is signaled by a jump. This abrupt change in stretch correspond to elongation if the tension is large enough to counterbalance the shortening effect of a positive torque acting on the two twisted ssDNA. As also predicted by the phase diagram in Fig.~\ref{ft}, the transition happens only at positive applied torque when tension exceeds the overstretching value, as DNA is then only stable under applied positive torque.  

\subsection{Anomalous overwinding}

Figures~\ref{twt},~\ref{dhg} show no anomalous overwinding under tension of the kind reported first by Gore et al.~\cite{Bust2}. As we discussed in Section I, this can be corrected by introducing somehow artificially a negative twist/elongation coupling in the purely mechanical energy of ssDNA, as other authors have done~\cite{Bust2,Gross}. If we do so---lack of structural rationale notwithstanding---we see that the fitting parameters can indeed be tweaked to reproduce the same phase diagrams and sharp transitions of Figures 2-6.

Fig.~\ref{twto} shows anomalous twisting under tension, for a choice  $o=-0.05$ of the parameter $o$. Note that $o$ as defined above in terms of the helical structure of dsDNA must be positive, and it controls the change in angular deviation of the helix under tension. However here we test the hypothesis that, because of  some hidden mechanism pertaining to the larger complexity of dsDNA (for instance the shrinking of its radius under tension~\cite{Bust2}), our framework could be simply modified  by introducing {\it ad hoc} a negative $o$, which can be done without dramatic consequences on our predictions so far. Then Fig. ~\ref{twto} shows profiles of supercoiling as a function of tension for DNA held under different torques, where we set at zero the angle corresponding to small tension. In the case of stretch under zero applied torque, we find results similar to those reported experimentally. As expected, the curve (black solid line) initially tracks $\tilde\Omega(f)-\Omega$, the purely mechanical tension-induced twist of dsDNA (which is positive when $o<0$), until tension approaches the overstretching transition and portions of ssDNA begin contributing negative supercoiling. 

Interestingly, and somehow more counterintuitively, Fig.~\ref{twto} shows that the effect increases when stretch is performed under a  negative torque, and then disappears for certain positive torques. This can be understood as consequence of the tension-increased torsional rigidity which further diminishes the effect of an applied negative torque as tension increases. At small tensions, most DNA is dsDNA, and  (\ref{Deltathetac0}) provides, for  stretching  under applied torque $\Gamma=\tau a$,  a twist given by
\begin{equation}
\Delta\theta_c(f)-\Delta\theta_c(0)=\tilde \Omega-\Omega+\tau\left[(\tilde \mu +\tilde \nu)^{-1}-( \mu + \nu)^{-1}\right].~
\end{equation}
Because the factor in the square parenthesis is always negative, a negative torque implies  a  twist that is larger than $\tilde \Omega-\Omega$ (dotted line in Fig.~\ref{twto}). Similarly a positive torque leads to a  twist that is lower than the purely tension-induced twist  $\tilde \Omega-\Omega$. And for stretch under zero applied torque, the experimental case as mentioned above, the twist simply tracks $\tilde \Omega-\Omega$. In all cases, the twist drops to large negative values as transition is approached and large fractions of ssDNA form. 

 This prediction could provide a purely mechanical experimental test that the anomalous overwinding originates in the non-trivial elastic structure of equilibrated dsDNA. And yet there might be alternative explanations: the overwinding could also be a thermodynamic effect, due to a decrement of ssDNA fraction due to the stabilizing effect of tension over flexural modes. It is true that our results do not seem to support such a scenario (Fig~\ref{twt}): the fraction of ssDNA at low tension and room temperature is already too tiny for its reduction to produce any substantial effect. Nonetheless we are disregarding nicks and defects in dsDNA, which might, under tension, increase the ssDNA fraction.

\section{Conclusions}

In this article we have introduced a simple  model, with a minimal number of parameters, to describe both ssDNA and dsDNA subjected to tension and torque, based on elementary considerations of symmetry and energetics. We have shown that  in a quadratic approximation in the torsional and flexural modes,  our treatment reduces to a modified PB model which now, unlike the original, incorporates the effect of fields, and carries information on angular and tensile degrees of freedom. 

We have used our approach to shown, analytically,  a range of predictions and phenomena for DNA thermomechanics inaccessible to previous models, or previously covered only partially and numerically.  We have compared our predictions with experimental results~\cite{Bryant, Mameren} and found strong agreement. {Some of these predictions, the phase diagrams and critical lines of Fig.~2-4  had been previously announced in a Letter~\cite{Nisoli11}. Others are new and required the introduction of flexural modes into the formalism. These are the probe/response profiles of Fig. 5-10. Our framework allows us to relate these temperature-induced or field-induced transitions to the density of ssDNA during transition, and  to find the linear dependences typical of first order transitions, witnessed experimentally~\cite{Mameren}. \\ \\
The introduction of flexural mode is intuitively necessary to explain extension and stretch under field. In this context our model proposes an explanation for the difference between angularly unconstrained and constrained over-stretching~\cite{Mameren}. \\ \\
Considering  Fig.~8, where we plot the computed torque exerted by the apparatus on the DNA molecule for angularly constrained traction-induced melting, we see that the system {\it situates itself  close to the critical line over a wide range of tension}. Indeed it seems that the molecule creates its own "intrinsic disorder" by progressively opening bases which, because of the lower torsional rigidity of ssDNA, can better counteract the unwinding effect of tension to maintain the fixed angle. While at small tension the torque applied by the apparatus to maintain the angular constraint grows linearly as one expects from a  purely mechanical description, as that curve approaches to the critical line, bases begin to open. Because ssDNA has much lower torsional rigidity, progressively much less compensating torque is exerted by the apparatus. Thus the system enforces the angular constraint by progressive and calibrated base opening, and maintains itself in such ``quasi-critical'' region for a broad range of tensions, until it reach the point of transition. 
It should be possible therefore to prepare the molecule in a region of high susceptibility just below transition, and that region corresponds to a distinctively large window in the controlling parameter (tension). This might be of consequence, should such a quasi-critical region of high susceptibility  be exploited for functionality, perhaps via enzymatic action or otherwise. As we wonder whether  ``biological systems are poised at criticality''~\cite{Mora} there is a  sense  that  the nano-machinery of life must exploit proximity to transitions to perform its tasks: total order is quiescent, requires large probes, and pertains to inanimate objects, complete disorder corresponds to death, but controlled disorder allows one to elicit a variety of responses and with smaller probes, something probably exploitable in replication, repair or transcription and in general for functionality.
}

In future work we will report on the effect of ionic strength. {As presaged in the text, variations in ionic strength can be reflected in the parameters within our model, especially $V_0$ and $r$. Electrostatic effects are in general non-local. However we will see in future work that at the zeroth approximation, for a DNA under tension and thus prevented from forming complex topologies,   this effect can be subsumed in a shift of the torque acting on the dsDNA, and thus on the equilibrium angular shift in the absence of external torque---a reflection of the fact that an electrically charged helix tends to unwind. This regime is accessible to our modeling framework precisely because it affords a coupling between torque and angular shifts. This approach to variations of ionic strength can be pursued even in the absence of applied tension, if the applied torque is small, because of dsDNA's long persistence length, and can thus explain denaturation induced by change of ionic strength.}
 
Finally, a transfer integral numerical study of the anharmonic version of~(\ref{Z}) is likely needed to refine our predictions at temperatures closer to denaturation, where the effect of nonlinearity in the stacking potential is important for the transition order and precursors~\cite{AB}.
\newline

\section{Acknowledgments} 

We are grateful to Z. Bryant (Stanford), J. Gore (MIT) and C. Bustamante (Berkeley) for stimulating conversation and for sharing experimental data, and to G. J. L. Wuite for stimulating conversation and for permission to reuse figures.

This work was carried out under the auspices of the National Nuclear Security Administration of the U.S. Department of Energy at Los Alamos National Laboratory under Contract No. DEAC52-06NA25396.


\begin{thebibliography}{}

\bibitem{Watson} J. D. Watson {\it et al.}, {\it Molecular Biology of the Gene} (Benjamin Cummings, New York, 2007).

\bibitem{Seeman2} N. C. Seeman, {\it J. Theor. Biol.} {\bf  99}, 237 (1982). 

\bibitem{Seeman} N. C. Seeman,  {\it Nature} {\bf 421}, 427 (2003). 

\bibitem{Winfree} E. Winfree, et al. {\it  Nature} {\bf 394} 539 (1998).

\bibitem{Yan} H. Yan et al. {\it Nature} {\bf 415} 62 (2002).

\bibitem{Pinheiro} A. V. Pinheiro et al. {\it Nature Nanotechnology} {\bf 6}, 763Ð772 (2011). 

\bibitem{Fu} T. J. Fu,  \& C. N. Seeman {\it Biochemistry} {\bf 32}, 3211 (1993). 

\bibitem{McFarlane} R. J. McFarlane et al. {\it Science} {\bf 334} 6053, 204 (2011).

\bibitem{Nykypanchuk} D, Nykypanchuk et al. {\it Nature} {\bf 451} (7178), 549-552	(2008).
	
\bibitem{DiMichele} L. Di Michele, E. Eiser {\it Physical Chemistry Chemical Physics} {\bf 15} (9), 3115-3129 (2013).

\bibitem{Rogers} W.B. Rogers  {\it PNAS} {\bf 108(38)} 15687-15692 (2011).	

\bibitem{Williams} K. A. Williams et al., {\it Nature} {\bf 420 }, 761 (2002).

\bibitem{Couet} J. Couet et al. {\it Angew. Chem. Int. Ed.} {\bf 44}, 3297 (2005)

\bibitem{Zheng} M. Zheng et al., {\it Nature Mater.} {\bf 2}, 338 (2003).

\bibitem{Zheng2} M. Zheng et al., {\it Science} {\bf 302}, 1545 (2003).

\bibitem{Gigliotti} B. Gigliotti et al. {\it Nano Lett.} {\bf 6}, 159 (2006)

\bibitem{Yarotski} D. A. Yarotski et al., {\it Nano Lett.} {\bf 9}, 12 (2008).

\bibitem{Johnson3} R. R. Johnson et al., {\it Nano Lett.} {\bf 8}, 69 (2008). 

\bibitem{Johnson} R. R. Johnson et al., {\it Nano Lett.} {\bf 9}, 537 (2009).

\bibitem{Johnson2} R. R. Johnson et al. {\it Small} {\bf 6}, 31 (2010).

\bibitem{Manohar} S. Manohar, T. Tang, and Anand Jagota, {\it J. Phys. Chem. C} {\bf 111}, 17835 (2007).

\bibitem{Kilina} S. Kilina et al., {\it Journal of Drug Delivery} Article Id 415621 (2011)

\bibitem{Gannon} C. J. Gannon et al., {\it Cancer} {\bf110}  2654 (2007).

\bibitem{Ma} J. Ma {\it et al.} {\bf Science} {\bf 340} 1580 (2013).

\bibitem{Hickenboth} C. R. Hickenboth et al. {\it Nature} {\bf 446} 423 (2007).

\bibitem{Sulc} P. Sulc, et al. {\it The Journal of chemical physics} {\bf 137} 135101, (2012).


\bibitem{Bust} C. Bustamante {\it et al.},  {\it Nature} {\bf 421}, 423 (2003). 

\bibitem{Strick} T. Strick,  et al. {\it Progress in Biophysics and Molecular Biology} {\bf 74} 115, (2000)

\bibitem{Bryant} Z. Bryant {\it et al.}, {\it Nature} {\bf 424}, 338 (2003).

\bibitem{Bust2} J. Gore {\it et al.}, {\it Nature} {\bf 442}, 836   (2006).

\bibitem{Mameren}  J. Van Mameren, {\it et al.},   {\it PNAS}  {\bf 106} 18231 (2009).

\bibitem{Smith96} S. B. Smith, Y. Cui, C. Bustamante 
 {\it Science} {\bf 271} 795 (1996)

\bibitem{LŽger} J. F. L\'eger {\it et al.},   {\it Phys. Rev. Lett.} {\bf 83}, 1066 (1999).

\bibitem{Tempestini} A. Tempestini, et al. {\it Nucleic acids research} {\bf 41}  2009 (2013).



\bibitem{Sarkar} A. Sarkar {\it et al.},   {\it Phys. Rev. E} {\bf 83}, 051903 (2003).

\bibitem{Marko} J. F Marko {\it Europhys. Lett.} {\bf 38}, 183 (1997).

\bibitem{Fye} R. M. Fye, C. J. Benham, {\it Phys Rev E} {\bf 59}, 3408 (1999).

\bibitem{Bouchiat} C. Bouchiat, M. M\'ezard {\it Eur. Phys. J. E}, {\bf 2}, 377 (2000)

\bibitem{Manghi} M. Manghi {\it et al.}, {\it J. Phys.: Condens. Matter} {\bf 21}, 034104 (2009).

\bibitem{Nisoli1} C. Nisoli {\it et al.},    {\it Phys. Rev. Lett.} {\bf 104}, 025503 (2010). 

\bibitem{Palmeri} J. Palmeri {\it et al.}. {\it Phys. Rev. Lett.} {\bf 99}, 088103 (2007). 


\bibitem{PB} M. Peyrard {\it et al.},    {\it Phys. Rev. Lett.} {\bf 62}, 2755 (1989).

\bibitem{Nisoli09} C. Nisoli {\it et al.} {\it Phys. Rev. Lett.} {\bf 102} (24), 245504 (2009).

\bibitem{PBD} T. Dauxois, M. Peyrard, {\it Phys. Rev. E} {\bf 51}, 4027 (1995).

\bibitem{Gross} P. Gross {\it et al.} {\it Nature Physics} {\bf 7} 731 (2011).

\bibitem{Dauxois} T. Dauxois, M. Peyrard, {\it Physics of Solitons} (Cambridge University Press, 2010).

\bibitem{Cocco} S. Cocco and R. Monasson, {\it Phys. Rev. Lett. } {\bf 83}, 5178 (1999).

\bibitem{Barbi} M. Barbi, S. Cocco, and M. Peyrard, {\it Phys. Lett. A} {\bf 253}, 358 (1999).

\bibitem{Nisoli11} C. Nisoli and A. R. Bishop, {\it Phys. Rev. Lett.}  {\bf 107}, 068102 (2011).

\bibitem{DiMichele2} L. DiMichele {\it et al} {\it JACS} {\bf 136(18)} 6538-6541 (2014).

\bibitem{Yan2} J. Yan, J. F. Marko, {\it Phys. Rev. Lett.} {\bf 93}, 108108 (2004).

\bibitem{Rapti} Z. Rapti et al. {\it Europhysi. Lett.} {\bf 74} 540 (2006).

\bibitem{Kleinert} H. Kleinert {\it Path Integrals in Quantum Mechanics, Statistics, Polymer Physics, and Financial Markets} (World Scientific, Singapore, 2006).


 
 \bibitem{Marko95} J. F. Marko; E. D. Siggia {\it Macromolecules} {\bf 28}  8759 (1995)
  
\bibitem{Odijk95} T. Odijk,  {\it Macromolecules} {\bf 28} 7016 (1995).

\bibitem{Wang97}  M. D. Wang,  H. Yin, R. Landick, J. Gelles and S. M. Block (1997), {\it Biophysical Journal} {\bf 72} 1335 (1997).

\bibitem{Landau} L. D. Landau, L. M. Lifshitz, {\it Quantum Mechanics} Butterworth-Heinemann (1976)


\bibitem{Saenger} W. Saenger, {\it Principles of Nucleic Acid Structure} (Springer-Verlag, New York, 1984).

\bibitem{Ha} H. Fu, {\it et al.},    {\it Nucleic Acids Res.} {\bf 38} 5594 (2010).


\bibitem{Hurst} L. D. Hurst and A. R. Merchant, {\it  Proceedings of the Royal Society of London. Series B: Biological Sciences} {\bf 268} (1466) 493, (2001).

\bibitem{Kashefi} K. KasheÞ and D. R. Lovley, {\it Science}, {\bf 301}, 934, (2003).

\bibitem{Kafri} Y Kafri at al, {\it Phys. Rev. Lett.} {\bf 85} 4988 (2000).



\bibitem{Romano} F. Romano et al. {\it Journ. Chem. Phys.}  {\bf 138}, 085101 (2013).

\bibitem{Cluzel} P. Cluzel et al. {\it  Science} {\bf 271}:792 (1996).

\bibitem{Wenner} J. R. Wenner et al {\it  Biophys J} {\bf 82} 3160, (2002).


\bibitem{Rye} H. S. Rye  et al. {\it Nucleic Acids Res} {\bf 20} 2803 (1992).

\bibitem{AB} N. K. Voulgarakis {\it et al.}, {\it Nano Lett.} {\bf 4}, 629   (2004).

\bibitem{Mora} T. Mora, W. Bialek {\it Journal of Statistical Physics} {\bf 144} (2), 268-302 (2011).


\end{thebibliography}
\end{document}